\documentclass[epj]{svjour}
\usepackage[dvipsnames]{xcolor}
\usepackage{amsmath,amssymb}
\usepackage{graphicx}
\usepackage{bm}
\usepackage{epstopdf}

\begin{document}
\title{Nuclear and Quark Matter at High Temperature}
\titlerunning{QM from HTFT}
\author{ 
Tam\'as S. Bir\'o\inst{1} 
\thanks{Supported by NKFIH/OTKA project No. 104238 and 104260. }
\and 
Antal Jakov\'ac\inst{2} 
\and
Zsolt Schram\inst{3}
}
\authorrunning{T.S.~Bir\'o \and A.~Jakov\'ac \and Z.~Schram}
\institute{%
H.A.S. Wigner Research Centre for Physics, Budapest, \email{Biro.Tamas@wigner.mta.hu}
\and %
Roland Eotvos University, Budapest, \email{jakovac@caesar.elte.hu}
\and %
Institute for Theoretical Physics, University of Debrecen, \email{schram@phys.unideb.hu}
}
\date{Version: \today / Received: date / Revised: another date}
\abstract{%
We review important ideas on nuclear and quark matter description on the basis
of high-temperature field theory concepts, like resummation, dimensional reduction,
interaction scale separation and spectral function modification in media.
Statistical and thermodynamical concepts are spotted in the light of these
methods concentrating on the -- partially still open -- problems of the hadronization
process.%
}
\maketitle

\renewcommand{\d}{\partial}
\newcommand{\ph}{\varphi}
\newcommand{\exv}[1]{\left\langle{\: #1 \:}\right\rangle}
\newcommand{\exvs}[1]{\langle{#1}\rangle}
\newcommand{\ep}{\varepsilon}
\newcommand{\GeV}{\textrm{~GeV~}}
\newcommand{\Tr}{\mathop{\textrm{Tr}}}
\newcommand{\pint}[2]{{\int\!\frac{d^{#1}#2}{(2\pi)^{#1}}\,}}
\newcommand{\MSbar}{{\ensuremath{\overline{\mathrm{MS}}}}}

\newcommand{\pd}[2]{{\frac{\partial #1}{\partial #2}}}
\newcommand{\be}{\begin{equation}}
\newcommand{\ee}[1]{\label{#1} \end{equation}}
\newcommand{\ba}{\begin{eqnarray}}
\newcommand{\ea}[1]{\label{#1} \end{eqnarray}}
\newcommand{\nl}{\nonumber \\}
\newcommand{\vs}{\vspace{5mm}}

\renewcommand{\c}[1]{{\cal{#1}}}
\renewcommand{\Im}{\, \mathfrak{Im}\, }
\renewcommand{\Re}{\, \mathfrak{Re}\, }

\renewcommand{\d}{{\, \mathrm{d}}}
\newcommand{\nn}{\nonumber\\}
\newcommand{\NL}{\nonumber \\}

\newcommand{\infi}{ \int_0^{\infty}\limits \,}
\newcommand{\ead}[1]{ \, \mathrm{e}^{#1} \,}

%
\section{Introduction}
%

At the roots of thermal field theory, 
back to the late 1960-s, field theory calculations and ideas applied to nuclear physics
were considered as ''exotic'' as the idea of using heavy atomic nuclei as projectile and target
in high energy accelerator experiments. In these heroic times the most prominent idea was to
experimentally produce and study {\em very hot nuclear matter}, whatever it shall 
be\cite{HISTORY,HISTORYB}.

Parallel to the achievements of QCD and the Standard Model of particle physics, the idea of a phase
transition from ''normal'' nuclear and hadronic matter to a quark-gluon plasma (QGP) have emerged
\cite{QGP,QGPB,QGPC}.
Transgressing the ideas of nuclear democracy \cite{NUCDEM,NUCDEMB}  
and an infinite tower of had\-ron\-ic resonances 
not allowing to exceed the Hagedorn temperature \cite{HAGEDORN}, 
the MIT bag model of hadrons based speculations
about a phase transition to a plasma of free colored charges, a QGP, became popular \cite{BAG,BAGB}.
This and the more and more progressing nuclear fluid treatment \cite{NUCFLUID,NUCFLUIDB}
at high bombarding energies
in the range of 1 GeV/nucleon and upwards in fixed target experiments let the hydrodynamical models
flourish. Since hydrodynamics relies only on the local conservation of energy, momenta and
eventually of a few more Noether currents, the only input needed to carry out such calculations
is an equation of state, a connection between local pressure and energy density.
In this way it provides a flexible framework to test underlying theories predicting
various equations of state \cite{MODERNHYDRO}.

In the forthcoming decades it has been gradually revealed, that neither the QGP, nor the transition
process is as simple as originally proposed. 
A remnant of confining forces, a long range correlation between colored particles, in some
respect reminding to (pre-)hadrons, in some other respect not being particle-like at all,
pollute the naive picture of a free QGP \cite{CONF,CONFB}. 
More devastatingly, the non-perturbative infrared
effects occur not only at low temperature, but with a low relative momentum between any pairs
of particles at all temperatures \cite{ALLCONF,ALLCONFB}. Also the color deconfinement phase transition, 
at the beginning surmised to be of first order with a huge latent heat density, proved
to be of a rather continuous transition with no uniquely fixed transition temperature point
by more recent lattice QCD calculations with dynamical light quarks \cite{LATTICE,LATTICEB}.
The ''exact'' transition temperature does not exist, only a position for a maximum in
one or another susceptibilities can be obtained. While a general lowering trend in the
{\em deconfinement temperature}, $T_c$, can be observed from $175$ MeV through $165$ MeV for a long time
and recently down to $158$ MeV,
the width of the transition zone is about $15-35$ MeV itself. Since
the transition is not of first or second order at vanishing baryochemical potential,
the ''correct and only'' order parameter cannot be 
identified \cite{FODORA,FODORB,FODORC,FODORD,WIDETRANS}.

There are furthermore doubts about the applicability of hydrodynamics \cite{ILLUSION} 
at the very early stage
of heavy ion collisions and at the final hadronization process, when the quarks and gluons
suddenly form hadrons. The details of the latter process are still unresolved; phenomenology
based fragmentation functions and modeling level string- or rope-decay scenarios are in 
use \cite{ROPEA,ROPEB,ROPEC,ROPED,ROPEE}.
For the early phase, when nevertheless most of the final state entropy is supposed to be
produced already, pictures utilizing the concept of coherent, nearly classical color fields
dominate, describing color rope formation and more recently a colored glass condensate 
(CGC) \cite{CGCA,CGCB,CGCC,CGCD}.

In this concise review we shall concentrate on selected issues related to applications
of models and achievements of high temperature field theory to nuclear physics, in particular
to relativistic heavy ion reactions. After a short review of the properties of quark matter
we deal with basic concepts of the hierarchy of scales and dimensional reduction.
Then considering the structure of the QGP we review the spectral function approach
and its main consequences for the medium properties, including the shear viscosity. 
This is followed by a review of special, nonlinear coherent states, showing a possibility
to produce negative binomial distribution of numbers in quantum states.
Finally a short conclusion section
rounds up this brief review with indications of some open problems in the field.

\section{Properties of Quark Matter}

Our picture about the properties of quark matter and the very definition of quark matter and 
quark--gluon plas\-ma (QGP)
underwent some changes in the passing decades. Starting with the picture of the plasma state
as a ''fourth phase'' beyond solid, liquid and gas, and responding to the idea of local liberation of
color charges, by now almost all quark or QCD-level descriptions, also that of a hadronic resonance
gas or string theory fitting numerical lattice QCD equation of state results, are considered
as dealing with ''quark matter''.
We have learned step by step (and by doing more and more precise ab initio numerical experiments
on bigger and bigger computer farms) that the QGP should have a very rich interaction structure.
Around the transition to color deconfinement in terms of temperature and chemical potentials, in a
grand canonical approach expanded in terms of the chemical potential to temperature ratio,
$\mu/T$, the real state of matter is far from
having free color charges, quarks and gluons, in a classical ideal gas based plasma.
Not only that the color freedom is only ''asymptotic'', being expressed only between pairs
having relative momenta sufficiently larger than a characteristic scale, whose estimates
range from $3T_c$ to $6T_c - 10 T_c$, but also hadron--like correlations survive well in the temperature
zone of $T_c - 4T_c$ according to modern lattice data.

Beyond heavy mesons, like the $c\overline{c}$ or $b\overline{b}$ 
system, also new, on the hadron level exotic
complexes, like glueballs, dibaryons, pentaquarks, etc. have been considered as playing
a crucial role in forming the rich structure of the realistic QGP near and above $T_c$.
In particular the $1/T^2$ fat tail of the interaction measure, $(e-3p)/T^4$, at high
temperature ($T \in [T_c,4T_c]$), that is so luring to be interpreted as a mass term, $m^2/T^4\sim 1/T^2$,
has been given special thoughts by several authors \cite{ALLCONFB,SEMIQGP1,SEMIQGP2,SEMIQGP3}.
Also the question of critical endpoint in the $T - \mu_B$ plane, signaling the border
between a first order color deconfinement phase transition and a continuous crossover
between hadronic resonance gas and QGP, has been studied in deep details relating different
susceptibilities to the quality of underlying ''freed'' color degrees of freedom
\cite{FLUCKOCH1,FLUCKOCH2,FLUCKOCH3,FLUCKOCH4,FLUCKOCH5,FLUKARS1,FLUKARS2,FLUKARS3,FLUKARS4}. 
Finally the problem of a quarkyonic phase, the expected structure
of quark matter at low temperature but high baryon density, and the coincidence or not coincidence 
of the color deconfinement transition with the chiral symmetry restoring phase transition
are debated since long.

Beyond the plethora of more or less arbitrary (but often analytically tractable) models of QGP,
the lattice regularized approach to solving QCD non-perturbatively by numerical strategies
proved to be the one, which has received the most credits and trust in the community.
Although it also has its limitations, e.g. it cannot deal with dynamical processes on the
quantum level in real time, for the statistical -- thermodynamical approach it delivers
very useful insights into a strongly coupled, complex structure of matter, also called
newly an sQGP. It also helped a lot to identify keynote field configurations, like the
magnetic monopoles and the instantons, which may characterize the main physical difference
between confined (hadronic) and deconfined (QGP) states of matter.

However, in particular the perturbative QCD dominated regime is hard to be reached by numerical
simulation. Although by some tricky methods quite a few authors \cite{FODOR100TC} squeezed out
results at temperatures as high as $10T_c - 100T_c$, the real perturbative behavior,
also approached by traditional perturbative QCD (pQCD), sets in only at unrealistic high
temperatures. Certainly one of the problems is, that thinking in terms of temperature,
$T\approx T_c$ represents an average energy per degree of freedom, while in an accelerator experiment
bringing heavy ions to collide the spread of the relative pair-momenta goes in the order of several
dozens or even hundreds $T_c$. Therefore any approach can make only a part of the true
behavior of the physical QGP available, and our complex picture has to be constructed
based on the mosaics we have puzzled out so far. 

High temperature field theory, based on resummation and renormalization techniques starting
with analytic, perturbative approach, is a very special theoretical tool for obtaining
a more intuitive picture about sQGP than only analyzing lattice QCD results. Finally,
probably a comparison of correlation functions and density matrix elements obtained in both
ways shall tell us new, hitherto unheard stories about the ''real nature'' of quark matter.


Finally, 
it can be enlightening to review briefly the thermodynamics of ideal gases polluted with
objects having less than 3-dimensional kinetic degrees of freedom, but carrying strong and
possibly long ranged correlations. The most famous such objects are strings and ropes;
they feature quasi 1-dimensional objects inside the plasma. The free energy density
of an ideal gas will then be additively modified by an energy contribution reflecting the
average length, $\exv{\ell} \sim n^{-1/3}$, by a string tension, $\sigma$ as
\be
 f^{{\rm string}} = \sigma n \, n^{-1/3},
\ee{STRINGFREE}
besides the trivial $f^{{\rm id}}$ contributions. Here we present the simplest,
most straightforward implementation of this idea; more details can be taken from 
\cite{STRINGY1,STRINGY2,STRINGY3,STRINGY4}.

The non-relativistic ideal, non-equilibrium chemical potential in the Boltzmann approximation is given by
\be
 \mu_{{\rm id}} = \pd{f^{{\rm id}}}{n} \: = \: T \, \log \frac{n}{n_{{\rm eq, id}}(T)}.
\ee{IDEALMU}
The ideal gas free energy density contribution is its integral over the density $n$, resulting
\be
 f^{{\rm id}} = nT \left(\log \frac{n}{n_{{\rm eq, id}}(T)} \, - \, 1 \right) + f_0(T).
\ee{IDEALF}
The total free energy density is given as
\be
 f(n,T) \: = \: f^{{\rm id}}(n.T) + \sigma n^{2/3},
\ee{TOTALF}
leading to a non-equilibrium chemical potential
\be
 \mu(n,T) \: = \: \pd{f}{n} \: = \: \mu^{{\rm id}} + \frac{2}{3} \sigma n^{-1/3}.
\ee{TOTALMU}
This sum of a rising and falling function of the density, $n$, can be equal to a foreseen constant
-- in this simple example zero -- only above a critical temperature. Below that temperature
the plasma with strings would not reach any finite equilibrium density; the system must
disintegrate to disconnected objects, e.g. to hadrons.

In terms of scaled quantities the non-equilibrium chemical potential is described by a function,
\be
 \frac{\mu}{T} \: = \: g\left(\frac{n}{n_{{\rm eq, id}}(T)} \right).
\ee{SCALEDMU}
The key function corresponding to our above model is given by
\be
 g(x) \: = \: \ln x \, + \, 3 \lambda \, x^{-1/3}
\ee{gFUNCTION}
with
\be
 \lambda \: = \: \frac{2}{9} \, \frac{\sigma}{Tn_{{\rm eq, id}}^{1/3}(T)}.
\ee{lambda}
This function has its minimum for $g^{\prime}(x_m)=1/x_m-\lambda x_m^{-4/3}=0$,
giving $x_m=\lambda^3$. The condition for having a stable equilibrium density for the
QGP then follows from $g(x_m)=3\left(\ln \lambda + 1 \right)\le 0$ and reads as
\be
 \frac{2}{9} \, \frac{\sigma}{Tn_{{\rm eq, id}}^{1/3}(T)} \le e^{-1}.
\ee{CRITICALSTRING}
\begin{figure}
\begin{center}
 \includegraphics[width=0.6\textwidth]{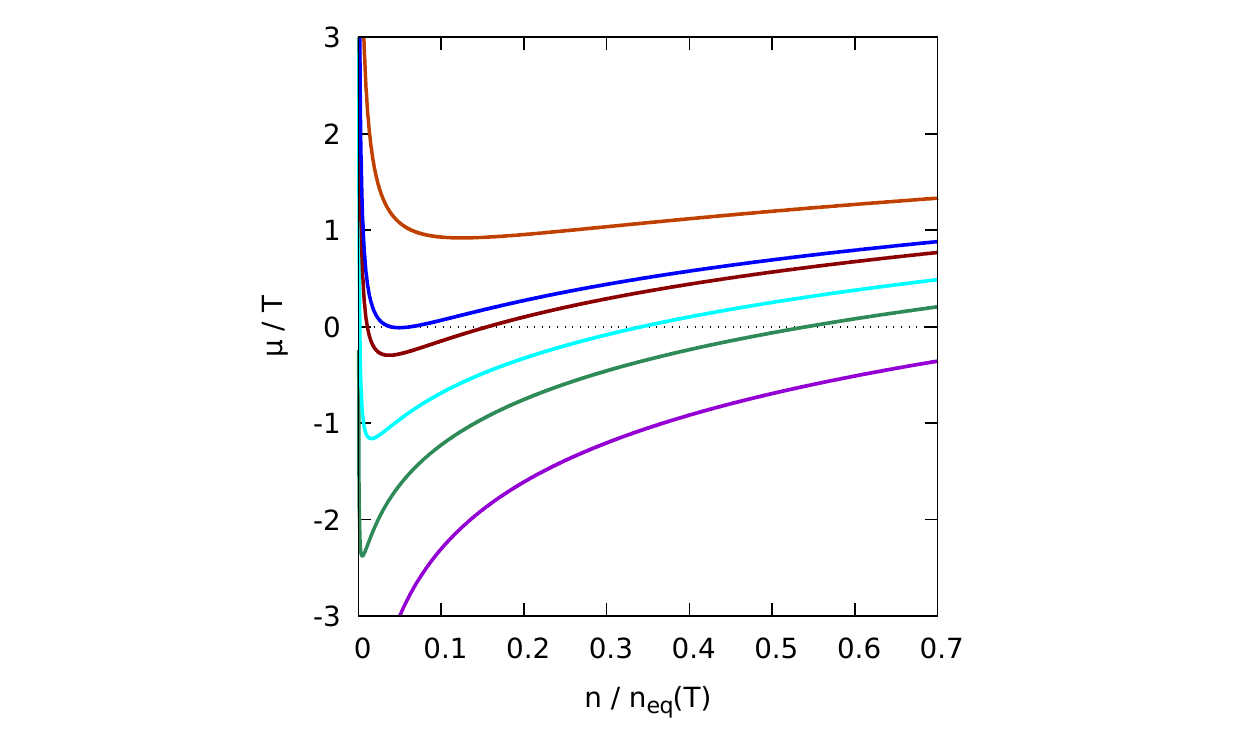}
\end{center}
\caption{ \label{FIG:STRINGYMU}
	The generic function $g(x)=  \ln x + \lambda x^{-1/3}$ for the scaled non-equilibrium chemical 
	potential, $\mu/T$, as a function of the scaled particle number density, $n/n_{{\rm eq, id}}(T)$.
	Various values of $\lambda$ from the bottom to the top line  are 
	$0, 0.5, 0.75, 1.0, 1.1, 1.5$.
}
\end{figure}
Interesting enough, that in the Boltzmann approximation, where $n_{{\rm eq, id}}(T)=T^3/\pi^2$
for each degree of freedom, and counting with the traditional $37$ effective light degrees
of freedom for a QGP we obtain the result that
\be
 T_{{\rm QGP}} \ge \frac{\sqrt{2}}{3} \, \frac{e^{1/2}\pi^{1/3}}{37^{1/6}} \, \sqrt{\sigma}
 \: \approx \: 0.623 \, \sqrt{\sigma}.
\ee{CRITICALT}
This result is near to the one obtained from early studies of the static quark -- antiquark
potential for the relation between the string tension and the
color deconfinement temperature in pure lattice gauge theories 
\cite{STRINGTENSION1,STRINGTENSION2,STRINGTENSION3,STRINGTENSION4}.	
This very simple-minded model can also be extended to finite baryochemical potentials, and does perform
appreciably \cite{STRINGY3}.	

\section{High-T effective field theory}

It is known since long, already from the perturbative QCD treatment of the quark-gluon
plasma (QGP) that interactions play a decisive role, and the description of equation
of state at high temperature cannot be based solely on the model of an ideal gas
of bare quarks and gluons. The more interesting that it can be and for a long time 
was being based on the ideal gas picture of quasiparticles, featuring the same number of
degrees of freedom as colored elementary quarks and gluons do. The most prominent effect
of interaction is concentrated to effective masses and its recursive effects on the
pressure, energy density and entropy density at a given temperature.

First experiences on nontrivial problems in the non-interacting quasiparticle treatment
arose from the study of the propagation of oscillatory excitations, so called plasmons,
in hot QGP: original calculations on the gluon damping coefficient, which determines
the speed of thermalization of a QGP, seemed to depend on the gauge fixing choice.
Even its sign was disputed in the beginning \cite{UKT1987,GAUGEDEP1,GAUGEDEP2,GAUGEDEP3}.

The solution was found by Rob Pisarski and Eric Braa\-ten with a resummation procedure
of the so called hard thermal loops (HTL-s) \cite{HTL1,HTL2,HTL3,HTL4,HTL5,HTL6,HTL7}.
The basis of this approach is a division of elementary quanta according to their
momenta: 'hard' are the hot thermal ones ($k \sim T$) and 'soft' are 
at momentum scales characteristic to the interaction ($k \sim gT$ to leading order).
Infinitely many Feynman graphs are grouped together so that the damping rate and
the effective mass (self energy in the infrared limit) can be calculated with methods
familiar from perturbative QCD. At high temperature the expansion according to the
coupling strength, $g$, and according to the number of loops in Feynman diagrams, $\hbar$,
is no more equivalent.

This, albeit is a big step forwards, does not solve alone all the problems.
Most prominently the static magnetic gluon mass is of order $g^2T$, occurs at a
'supersoft' scale, and cannot be generated by HTL resummation techniques alone.
One considers e.g. a dilute magnetic monopole gas, whose density is proportional
to $n \sim m^3\sim (g^2T)^3$, making a contribution to pressure and energy density at the level of
$p \sim nT \sim g^6T^4$. In the perturbative QCD approach this term is related to an infrared
divergence \cite{LINDE,INFRA}, and as such it is independent of UV renormalization schemes.
The magnetic gluon mass of order $g^2T$ seems to be of genuine nonperturbative
origin \cite{MAGMASS,MAGMASS2}. 
It plays a role also in the calculation of other physically relevant
quantities, like shear viscosity. Lattice QCD calculations on the other
hand obtained this static magnetic gluon mass via observing a reduced dimensional
string tension for space--space like Wilson loops as well as hunting for
magnetic monopole looking configurations during the Monte-Carlo integration
\cite{MONO1,MONO2,MONO3,MONO4,MONO5,MONO6}.

Basic formulas of high-temperature field theory make it possible to obtain
order of magnitude estimates by assuming different dominant gluon field
configurations, which contribute to the Euclidean path integral
integrating the factors $\exp({iS/\hbar})$ with the action
\be
 S_4 = i \int_0^{\hbar \beta}\limits\!d\tau \, \int\!d^3x \,
 \frac{1}{2} \left(E_{ia}^2+B_{ia}^2 \right).
\ee{S4}
Here the chromoelectric field is related to the vector potential via
the Euclidean time derivative, $E_{ia}=-\partial A_{ia}/\partial\tau$.
The quantum theoretical path integral,
\be
Z = {\rm Tr} \left(e^{\frac{i}{\hbar}S_4} \right) \: = \: 
{\rm Tr} \left( e^{- \frac{1}{\hbar} \int_0^{\beta\hbar}\limits H d\tau} \right),
\ee{Z4}
is carried out for (with their gauge equivalent) $\tau$-periodic $A_{ia}$ fields
with period $\beta\hbar$, and can therefore be reduced to a 3-dimensional
partition sum at high temperature (small $\beta\hbar$):
\be
Z_3 \: = \: {\rm Tr} \left( e^{-\beta H}\right)
\ee{Z3}
with
\be
H_3 \: = \: \frac{1}{2} \int\!d^3x \, \left(E_{ia}^2 + B_{ia}^2 \right).
\ee{HAM3}
In obtaining this result one assumes a constant $H(\tau)$ function in the narrow
interval $(0,\beta\hbar)$. This is relevant in the study of the infrared
behavior of the full, interacting theory.

For the sake of simplicity let us consider pure Yang-Mills theory (i.e. QCD without quarks)
for a while. The path integral trace is over vector potential configurations,
these can be re-scaled by the interaction strength: $gA \rightarrow A$ transformation
leads to an effective, reduced 3-dimensional action with an effective coupling
of the static magnetic mass
\be
 S_3 \: = \: \frac{1}{g^2T} \, \int\!d^3x \, \frac{1}{2} B_{ia}^2,
\ee{S3}
revealing the sought finite temperature partition sum as
\be
Z_3 \: = \: {\rm Tr} \left( e^{-S_3} \right).
\ee{Z3P}
Since this formula does not contain the Planck constant any more, we may confirm
that the chromo-magnetostatic features of QGP can be estimated by purely classical
field theory means. At the same time they are genuinely non-perturbative.

In the followings we review a few assumed gluon field configurations and investigate
the corresponding mass and density scales of gluons, in the original setting, before
re-scaling the vector potential with $g$.
As a starting point we have to relate the magnitudes of the vector potential
and that of the chromoelectric fields. We do this remembering that they are 
represented by canonically conjugate operators, satisfying
\be
\left[E_{ia}(\vec{x}), \: A_{jb}(\vec{x}) \right] \: = \: \frac{\hbar}{i}
\, \delta_{ij} \, \delta_{ab} \, \delta(\vec{x}-\vec{y}).
\ee{EACOM}
Looking for quantum states possibly near to classical fields one singles out
coherent states, where the Heisenberg uncertainty between the canonical operators
is minimal. \newline Henceforth we use the intuitive estimate
\be
 E \cdot A \sim  \hbar / L^3,
\ee{BASIC}
assuming a quantization box of length $L$. We classify the gluon field configurations
according to the magnitude of the vector potential and distinguish the following
three fiducial classes:
\begin{enumerate}

\item The vector potential is large, of classical order (independent of $\hbar$): 
$A \sim 1/gL$.
In this case $E \sim g \hbar/L^2$ and the magnetic field strength
becomes $B \sim A/L + g A^2 \sim 1/gL^2$, also classical. It receives Abelian and non-Abelian
contributions in equal magnitude.
The field energy,
\be
H = \frac{1}{2} \int\!d^3x \, (E^2+B^2) \: \sim \: 
\frac{1}{g^2L} \, \left( \, 1 + {\cal O}\left((g^2\hbar)^2\right) \, \right),
\ee{C1H}
is also classical and dominated by the magnetic contribution for $g^2\hbar \ll 1$.
Equating this value with the thermal gluon energy, $H \sim T$,
we obtain the relation $1/L \sim g^2T$, i.e. the supersoft magnetic scale
determines these configurations. The gluon density is estimated as
being $n \sim 1/L^3 \sim (g^2T)^3$ and the magnetic screening mass,
the gluon self energy in the infrared limit, is estimated from $m^2A^2 \sim g^2A^4$:
\be
 m^2 \sim  g^2A^2 \: \sim \: (g^2T)^2.
\ee{M2CLASS1}
This tour de force in estimates ends up with the mass $m \sim g^2T$.

\item
The vector potential and the electric field strength share the quantum order but 
they are independent of the coupling, $g$. In this case one typically deals with
configurations of $A \sim \sqrt{\hbar}/L$ and $E \sim \sqrt{\hbar}/L^2$.
The magnetic field is Abelian dominated,
$B \sim A/L + gA^2 \sim \sqrt{\hbar}/L^2 + g\hbar/L^2$.
The dominant chromomagnetic field is of the same magnitude as the chromoelectric one.
This describes a thermal state with equipartition and the energy density
\be
\varepsilon \sim \frac{T}{L^3} \: \sim \: \frac{\hbar}{L^4}.
\ee{HCLASS2}
The characteristic scale from this is obtained as the thermal wavelength,
$L \sim \hbar/T$, and the effective screening mass becomes
\be
m^2 \sim  g^2A^2 \: \sim \: \frac{g^2T^2}{\hbar} .
\ee{M2CLASS2}
This Debye screening mass is of the order $m\sim gT/\sqrt{\hbar}$.

\item
In principle a third class of configurations exists dominated by the classical chromoelectric field
on the account of a vector potential of highly quantum order:
$E \sim 1/gL^2$ and $A \sim g\hbar/L$. Physically this corresponds to the string picture
and implies $E \gg A/L$ for $g^2\hbar \ll 1$. The Abelian part of the chromomagnetic field,
$B_{Abel} \sim A/L \sim g\hbar/L^2 \sim g^2\hbar E$ is then smaller than the chromoelectric
field, while the non-Abelian contribution, $B_{non-Abel} \sim gA^2 \sim g^3\hbar^2/L^2$
is even smaller, negligible in the semiclassical weak coupling approach.
It is interesting that the thermal energy, dominated by the classical chromoelectric field,
\be
\varepsilon \: \sim \: \frac{T}{L^3} \: \sim \: \frac{1}{g^2L^4},
\ee{EPSI3}
again delivers a characteristic length scale of $L \sim 1/g^2T$. The screening mass
effect, however, in this case is very small and of highly quantum nature:
\be
 m^2 \sim g^2A^2 \: \sim \: \frac{g^4\hbar^2}{L^2},
\ee{M2CLASS3}
delivering at the end a mass scale of $m \sim g^2\hbar \cdot g^2 T$.
\end{enumerate}

Considering quasiparticles their mass is defined by the dispersion relation reflecting
the Schwinger-Dyson equation with a general, complex self-energy
\be
 \omega^2 - \vec{k}^2 - \Sigma(\omega,\vec{k}) \: = \: 0.
\ee{QDISPERS}
Interacting with a medium during propagation is included in the general self-energy term, 
$\Sigma$. The resolution of eq.(\ref{QDISPERS}) for $\omega$ can also deliver complex
values, the imaginary part of the frequency signalizes the so called plasmon damping.
In the infrared limit, $|\vec{k}|\to 0$ the general frequency is $\omega+i\gamma$,
satisfying
\be
 \left( \omega+i\gamma\right)^2 \: = \: \Sigma(\omega+i\gamma,0).
\ee{IRDISPERS}
In the weak damping limit, $\gamma \ll \omega$ the imaginary part of this equation defines
\be
\gamma \: \approx \: \frac{1}{2\omega} \, \Im \Sigma(\omega,0),
\ee{WEAKDAMP}
and the real part constitutes a mass gap equation
\be
\omega^2 \: \approx \: \Re \Sigma(\omega,0).
\ee{MASSGAP}
To leading order in the perturbative expansion $\Re \Sigma(\omega,0)=(m/\hbar)^2$ with
$m$ being the mass scale derived above. With $\omega\sim T/\hbar$ taken as hard thermal,
the weak damping constant becomes $\gamma \sim m^2/\hbar T$. This is perturbatively
the largest in the second class, $\gamma_2 \sim g^2T/\hbar$.

\section{Internal Structure of QGP}


The fact that QCD is a strongly interacting theory changes several
concepts originally stemming from the free particle world. 

\subsection{Particles in strongly interacting system}

In an interacting field theory the notion of a ``particle'' needs
careful definition. The problem is that the concept of a ``particle'' is associated
with \emph{free field theory}; but, in fact,
there are various definitions that refer to the same physical phenomenon in
non-interacting theories, any of them being appropriate to describe a free
particle:
\begin{itemize}
\item In free theory there exists a conserved particle number operator
  $\hat N$ that also commutes with the momentum operator, too. The
  common eigenvectors of the energy, momentum and particle number in
  the $N=1$ sector are the free particles. The $N>1$ sector consist of
  direct products of one-particle states; the direct sum of all
  $N$-particle sectors provides the Fock-space construction.
\item The energy $E$ and momentum $k$ of a free one-particle state is
  connected by the dispersion relation $E=E(\bm k)$. Therefore the
  spectrum of the one-particle sector consists of a single energy
  level, let us denote it $\left|E,\bm k\right\rangle$. The spectral
  density of this sector is therefore a single Dirac-delta. To measure
  the spectral density we can use any operator $\hat\Phi$ that posesses
  only a one-particle form factor, ie. $\langle E,\bm
  k|\hat\Phi|0\rangle\neq0$, but $\langle E,\bm k; E',\bm
  k'|\hat\Phi|0\rangle=0$. Then we define
  \begin{equation}
    \varrho(t)=\langle 0|[\hat\Phi(t),\hat\Phi(0)]_{\pm}|0\rangle,
  \end{equation}
  where $\pm$ refers to the commutator/anticommutator, depending on
  the bosonic/fermionic nature of the particles. In Fourier space this
  definition is equivalent with a weighted spectral density
  \begin{equation}
    \varrho(\omega) = \sum_{E,\bm k} \left|\langle
        E,\bm k|\hat\Phi|0\rangle\right|^2 2\pi\delta(\omega-E(\bm k)).
  \end{equation}
  In a relativistic field theory, using the fundamental field as a
  measurement operator we have the spectral function
  \begin{equation}
    \label{eq:freespect}
    \varrho(\omega,\bm k)=\frac{2\pi}{2E(\bm k)}\left[
      \delta(\omega-E(\bm k)) -\delta(\omega+E(\bm k))\right].
  \end{equation}
  This satisfies the sum rule
  \begin{equation}
    \label{eq:sumrule}
    \int\frac{d\omega}{2\pi} \omega\varrho(\omega,\bm k) =1,
  \end{equation}
\item The spectral function remains unchanged at finite temperature, so
  a particle at finite temperature is the same object as a zero temperature
  particle.
\item As a consequence the wave function of the free particle is
  $\Psi(t,\bm x) \sim e^{-iEt+i\bm{kx}}$, an infinite extension
  plane-wave, with $|\Psi|^2=1$ uniform probability density.
\item The linear response to a disturbance leads to the linear
  response function, or retarded Green's function. The retarded
  Green's function reads as
  \begin{equation}
    G_{ret}(\omega,\bm k) = \frac1{(\omega+i\varepsilon)^2-E^2(\bm k)}
  \end{equation}
 in relativistic systems. This form is preserved at finite temperature.
\end{itemize}
In free theory these all are consequences of each other, therefore we
unintentionally identify these concepts, and when we tell
``particle'', it means all of these at the same time.

However, in an interacting model all of these concepts yield different
results, and so we have to release the identification of the above
concepts.
\begin{itemize}
\item In a general field theory the number of conserved quantities is
much smaller than the number of state labels (types of quantizable physical degrees of freedom);
the only exceptions are integrable systems. 
In particular the particle number operator does not exists any more.
\item We can measure the spectral function in the same way as we did
  in the free case. But, because of the interactions, the spectrum of
  the free one-particle states will be mixed with the spectrum of the
  higher particle number states. These will provide a continuum
  contribution besides the free particle state. Since the spectrum is
  subject of a sum rule, cf. \eqref{eq:sumrule}, the height of the
  Dirac-delta peak can not be the free one, it receives a
  multiplicative correction $Z$ (wave function renormalization).
\item The spectrum is more complicated at finite temperature or at
  finite chemical potential: there the spectral function is nonzero
  for all frequencies (with the sole exception $\omega=0$), as a
  consequence of the scattering on particles in the
  environment. This broadens the Dirac-delta
  particle peak, resulting in a Lorentzian curve. Such an excitation is
  called \emph{quasiparticle}. It, as opposed to the free case, does
  not represent a single energy level, but a collection of energy
  eigenstates. Here other excited/ground states can appear too,
  and the continuum is always present. A typical finite temperature
  spectrum can be seen on Fig.~\ref{fig:toyspectrum}.
  \begin{figure}[htbp]
    \centering
    \includegraphics[height=4cm]{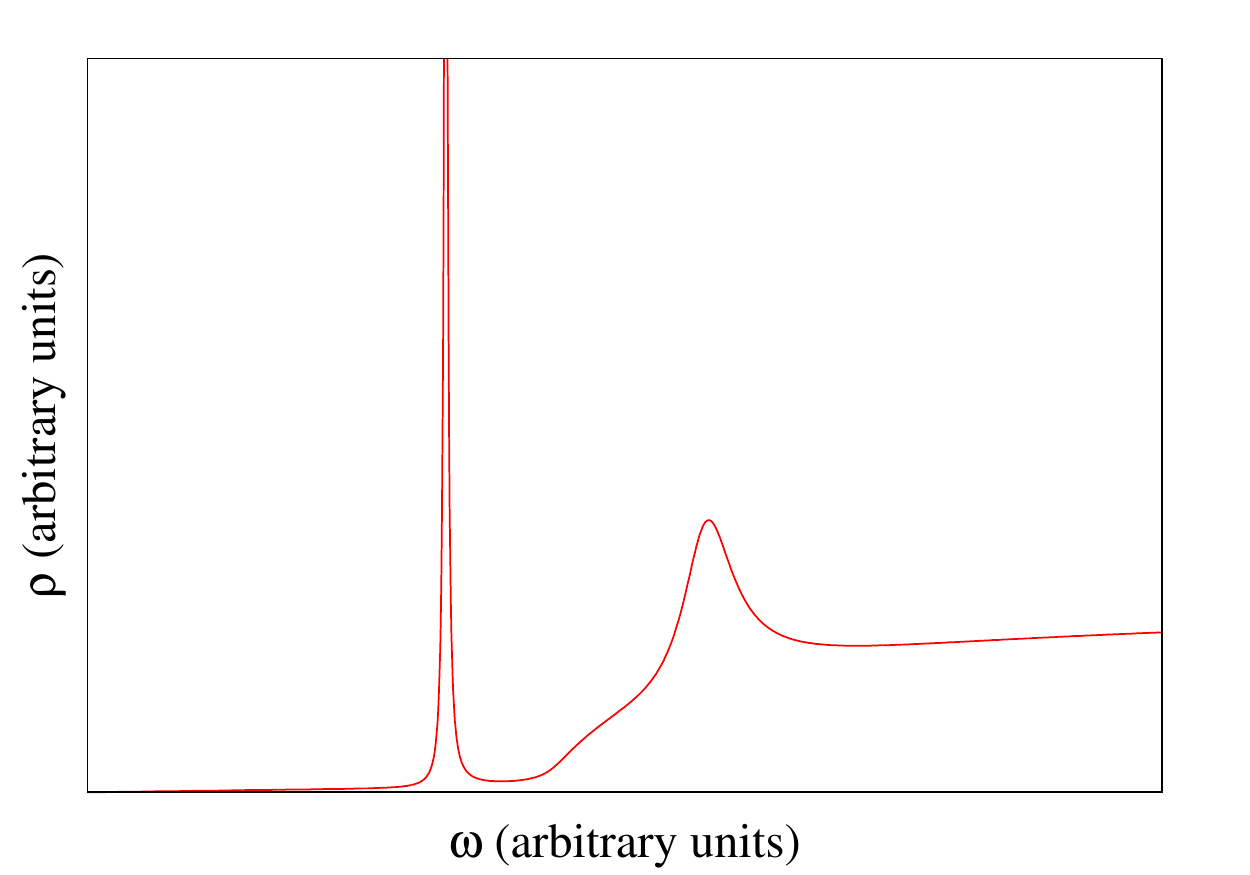}
    \caption{Typical spectrum in an interacting theory. The original
      particle peak is broadened (finite lifetime or finite coherence
      length), other peaks can appear (excited and bound states),
      moreover we always have a multi-particle continuum contribution.}
    \label{fig:toyspectrum}
  \end{figure} 
\item The linear response function at zero temperature still contains
  contribution from the Dirac-delta peak. For long times we obtain
  \begin{equation}
    G_{ret}(t,\bm k) = Ze^{-iE(\bm k)t} + C t^{-3/2} e^{-iE_{thr}(\bm
      k)t}+\dots.
  \end{equation}
  For long times the second part dies out, leaving a free particle
  like propagation: these are the \emph{asymptotic states}.
\item In numerous cases, however, there are no asymptotic states: if the
  particle mixes with zero mass particles (all charged
  particles do that), or the particle is not stable, or we are at nonzero
  temperature; practically in all realistic cases. The Lorentzian
  quasiparticle peak and the continuum part of the spectral function
  yield the retarded propagator
  \begin{equation}
    G_{ret}(t,\bm k) = Ze^{-iE(\bm k)t-\gamma_{\bm k}t} + C t^{-3/2} e^{-iE_{thr}(\bm
      k)t-\Gamma_{\bm k}t}+\dots.
  \end{equation}
where $\gamma_{\bm k}$ is the half-width of the Lorentzian peak, and
$\Gamma_{\bm k}$ is some parameter determining the smoothing of
the spectrum near the threshold. The quasiparticles for long times
decay exponentially\footnote{We must emphasize, however, that this
is true for long times only, for short times a power-law like
damping is also possible.}.  If $\gamma_{\bm k}\ll\Gamma_{\bm k}$,
for long times we can observe a fading quasiparticle response, in
the reverse case, $\gamma_{\bm k}\gg \Gamma_{\bm k}$, the long time behavior 
of the system is not particle-like at all.
\end{itemize}
Having said all these we see that the particle concept becomes a
dangerous ground, we must be very precise on what we are talking
about. For example stating that the particles have temperature
dependent mass is sensible only in the quasiparticle-sense: free
particles are eigenstates of the Hamiltonian, they cannot have a mass
changing with the temperature. The quasiparticles, on the other hand,
are collections of energy eigenstates, and the coefficients of the
combination can change with $T$. Therefore the position and the width
of the quasiparticle peak can also change with the temperature.

\subsection{Particles, spectral function and thermodynamics}

There is still a possible definition for a particle through
thermodynamics: in free theory each particle species represents a
thermodynamic degree of freedom. Does it remain true in the
interacting case, i.e. are also the nonperturbative particle species
thermodynamic degrees of freedom? Let us seek an answer to this
question by utilizing spectral properties.

The pressure of the free gas of different species is the sum of
partial pressures $P= \sum_n P_n$ with
\begin{equation}
  \label{eq:freeP}
  P_n=\mp\frac{T^4}{2\pi^2}\int\limits_0^\infty\!dx\,x^2\ln(1\mp
  e^{-\sqrt{x^2+(\beta m_n)^2}}),
\end{equation}
where $m_n$ is the mass of the $n$th particle, and $\mp$ refers to
the bosonic/fermionic case, respectively. In particular the
pressure at large temperature, in the Stefan-Boltzmann
limit, $T\gg m$, reads as
\begin{equation}
  \label{eq:SBlimit}
  P_{free} = \frac{\pi^2}{90}\left(N_b+\frac78 N_f\right),
\end{equation}
where $N_b$ is the number of bosonic, $N_f$ the number of fermionic
species. All fundamental free particles therefore contribute
independently to the pressure.

We do not need to have, however, fundamental particles to obtain thermodynamical
degrees of freedom.  In the case of well separated quasiparticle
excitations, that can be characterized by a phase shift $\delta(\omega)$
of a scattered particle centered at $\omega=E(\bm k)$, the
Beth-Uhlenbeck formula \cite{Landau} states
\begin{equation}
  \label{eq:BethUhlenbeck}
  \delta Z \sim \int\limits_0^\infty
  \frac{d\omega}{2\pi}\frac{\partial\delta}{\partial\omega}
  e^{-\beta\omega} \sim e^{-\beta E(\bm k)}.
\end{equation}
Therefore all excitations contribute to the partition function exactly
like a stable particle, irrespective whether it is a fundamental
particle, or a bound state with internal structure and motion.

This picture leads to the Hadron Resonance Gas (HRG) description of
the QCD plasma \cite{HRG0}. Here all the possible hadrons, measured
and identified at zero temperature, contribute to the thermal ensemble
in the same way, irrespective of their width. The resulting
pressure is in fact in a very good agreement with the pressure
measured in MC simulations in the hadronic phase
\cite{Andronic:2003,Karschetal,Huovinen:2009yb,Borsanyi:2010cj,Bazavov:2013dta}.

It fails, however, badly in the quark-gluon phase, about $T> 150$ MeV
\cite{Bazavov:2013dta,FLUKARS4}. In fact, as it was first
pointed out by Hagedorn, the HRG pressure would be divergent, if all
hadrons were taken into account. This is due to the fact that the
density of hadron states grows exponentially with the mass:
$\rho(m)\sim m^\alpha e^{\beta_H m}$, with $T_H=1/\beta_H$ 
Hagedorn temperature, and an appropriate power $\alpha$ (e.g.
$\alpha=5/2$). Then the pressure of all hadrons, written up as an
integral for the mass density diverges,
\begin{equation}
  P \sim \int\!\!dm \, \rho(m) \, e^{-\beta m} \: \to \:
  \infty, \qquad\mathrm{for}\qquad T>T_H.
\end{equation}

If we do not take into account all hadrons, just those that are listed
in the Particle Data Book \cite{PDG}, or the hadrons below, say 3 GeV,
then the pressure will not diverge, but still overshoots the pressure
of the quark gluon plasma. This is a conceptual problem:
at all temperatures the system in equilibrium 
realizes that phase where the grand-canonical thermodynamical
potential, in this case $PV$, is the largest. 
The quark gluon plasma with 8 gluons and $N_f$ quarks
represents a system with $16+10.5 N_f$ bosonic degrees of freedom (all
fermions have 4 Lorentz-components, 3 colors, and the factor $7/8$ compared
to the bosonic contribution). If we would count only the stable
hadrons, i.e. pions and nucleons, as hadronic degrees of freedom, we would
obtain $10$ bosonic degrees of freedom, and so the QGP would have a
larger number of degrees of freedom, which explains, why there is a
phase transition to the QGP phase. But if all the Particle Data Book hadrons are taken
into account, this highly exceeds the QGP number of degrees of
freedom, and we do not understand, why a phase transition occurs at
all?  We must emphasize that the argument that the hadrons are not
valid degrees of freedom in the QGP phase is not applicable, since the
hadron phase represents a higher entropy state of matter, and so it
forbids the change to the QGP phase.

So we are faced with the situation, where we can describe the pressure
of the strongly interacting plasma below $\sim150$ MeV (HRG), and
above $300$ MeV (QGP), but we do not understand why there is a
phase transition, and we do not understand the pressure in the
intermediate temperature range. What happens with the hadrons between
$150\,\mathrm{MeV}< T<300\,\mathrm{MeV}$? The bound states must
somehow disappear from the system as we rise the temperature,
physically the hadrons must \emph{melt} away. 

How is this melting related to the Beth-Uhlenbeck formula, according to that
every quasiparticle resonance corresponds to a single
thermodynamical degree of freedom? We should note that in the
derivation of the result one must assume that the quasiparticles are
independent, in the sense that all can be treated as separate
Breit-Wigner peaks. This assumption, however, fails when we 
consider a system where the quasiparticle peaks occur
densely, or if a multiparticle background is present. In such cases
the quantum mechanical treatment of the quasiparticle peak contributions to
the $S$ matrix have complex coefficients \cite{Scon,Svec}. The
unitarity of the $S$ matrix poses constraints among these
coefficients: in this way the pole contributions are no more
independent.

In field theory we can describe this process by observing the hadronic
spectral functions
\cite{Jakovac:2012tn,Jakovac:2013iua,ALLCONFB}. A mathematically
similar description can be obtained using the Mott-transition analogy
\cite{Turko:2011gw,Benic:2013tga,Blaschke:2015nma}. The hadronic
spectral functions, as all spectral functions, consist of a
quasiparticle peak and a continuum part. The weight of these parts,
however, changes with the temperature. At small temperatures the
quasiparticle peak is pronounced, it dominates the thermodynamics, and
the gas of hadrons behaves as a gas of almost free particles. At high
temperatures, however, the quasiparticle peak merges with the
continuum, and the ``particle'' nature of the hadronic channel ceases
to be true. This is accompanied by a drastic reduction of the partial
pressure in this channel.

To set up a field theoretical model we construct a quadratic theory
with the same statistical property (boson/fermion) and the same
spectral function as the studied channel. For a scalar field it means
that we write up the Lagrangian as
\begin{equation}
  \label{eq:effL}
  {\cal L} \: =  \:  \frac12 \Phi \, {\cal K}(i\partial) \, \Phi,
\end{equation}
with some kernel ${\cal K}$. The kernel and the retarded Green's
function are related as $G_{ret}(p) ={\cal K}^{-1}(p_0+i\varepsilon,\bm p)$,
while the retarded Green's function and the spectral function can be
expressed from each other
\begin{equation}
  \label{eq:rhoGret}
  G_{ret}(p) \: = \: \int\limits_{-\infty}^\infty\! \frac{d\omega}{2\pi}
  \frac{\varrho(\omega,\bm p)}{p_0-\omega+i\varepsilon},\quad
  \varrho(p) =\mathop{\mathrm{Disc}}_{p_0} iG_{ret}(p).
\end{equation}
Since the theory described by the effective Lagrangian \eqref{eq:effL}
is quadratic, and so it is solvable; but its spectrum does not consists
of free particles. To determine thermodynamics one has to start from a
microscopically measurable quantity, which most conveniently can be
chosen the energy density, ie. the expectation value of the 00
component of the energy-momentum tensor $T_{00}$. Although we have a
quadratic, model, the energy-momentum tensor is not simple, due to the
nonlocal nature of the kernel. The divergence of the energy-momentum
tensor can be determined from the variation of the action with respect
to a space-time translation $a_\mu$ \cite{Collins}
\begin{equation}
  \partial^\mu T_{\mu\nu} = \frac{\delta S}{\delta a^\mu}.
\end{equation}
This leads to \cite{Jakovac:2012tn} \newcommand{\K}{{\cal K}}
\begin{equation}
  T_{\mu\nu}(x) = \frac12 \Phi(x)\,D_{\mu\nu}\K(i\partial)\, \Phi(x)
\end{equation}
where
\begin{equation}
  D_{\mu\nu}\K(i\partial) = \left[\frac{\partial\K(p)} {\partial p^\mu} \biggr|_{p\to
      i\partial}\right]_{sym}\!\!\! i\partial_\nu - g_{\mu\nu}\K(i\partial),
\end{equation}
and the symmetrized derivative is defined as
\begin{equation}
  f(x) [(i\partial)^n]_{sym} g(x) = \frac1{n+1}\sum\limits_{a=0}^{n} [(-i\partial)^a
  f(x) ] [(i\partial)^{n-a} g(x)].
\end{equation}
Once we know $T_{\mu\nu}$, wa can take its expectation value in
equilibrium. We can use KMS relation to write
\begin{equation}
  \left\langle\Phi(x)\Phi(y)\right\rangle= \int\!\frac{d^4p}{(2\pi)^4}
  e^{-ip(x-y)} \left(\frac12+n(p_0)\right) \varrho(p).
\end{equation}
Finally we renormalize the expressions and express pressure through
thermodynamical relations. The result reads
\cite{Jakovac:2012tn,Jakovac:2013iua,ALLCONFB}
\begin{equation}
  \label{eq:exP}
  P = \mp T \int\!\frac{d^4p}{(2\pi)^4} \,\Theta(p_0) \frac{\partial{\cal
      K}}{\partial p_0} \varrho(p) \ln(1\mp e^{-\beta p_0}).
\end{equation}
We should note that the pressure does not depend on the normalization
of the spectral function, since the kernel is inversely proportional
to this normalization factor (cf. \eqref{eq:rhoGret}).

\begin{figure}[htbp]
  \centering
  \hspace*{-3em}\includegraphics[height=4cm]{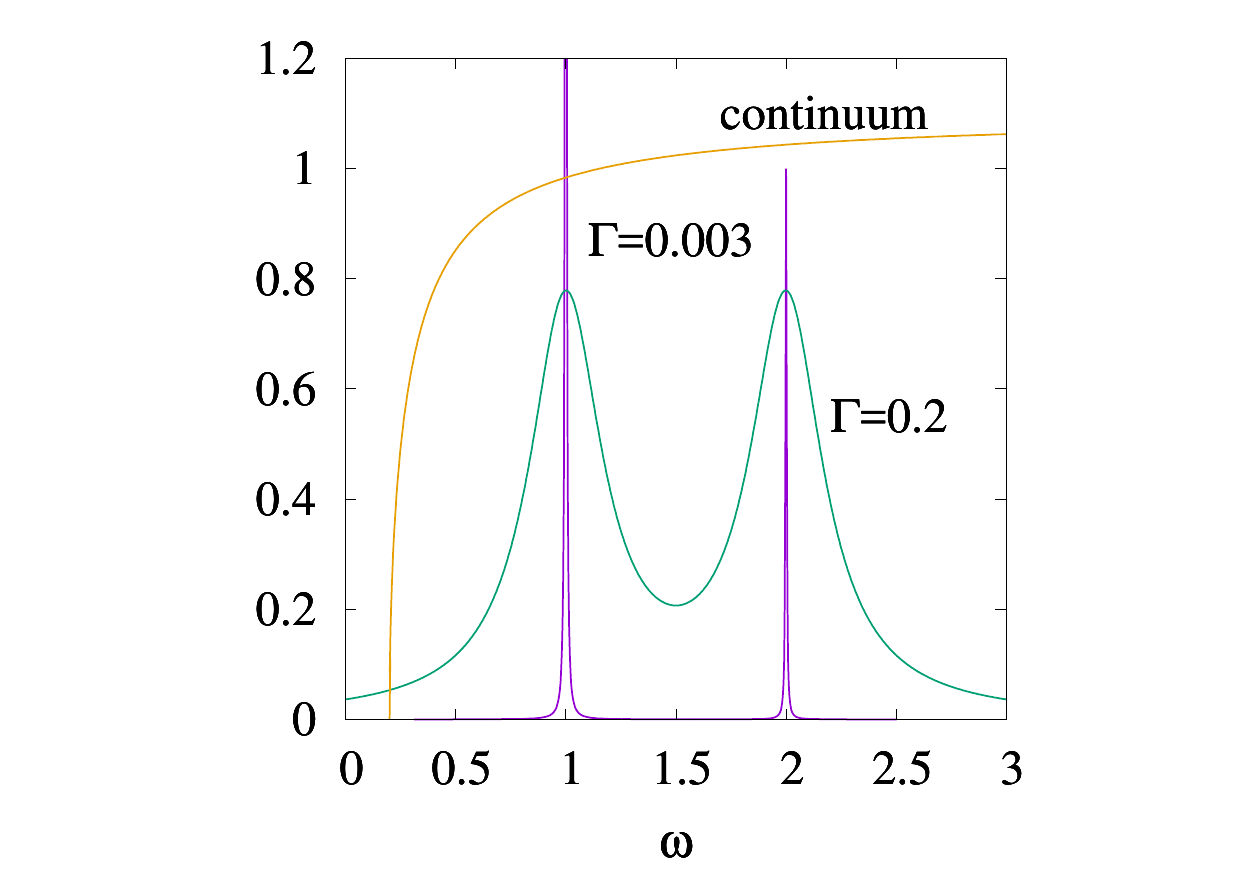}\hspace*{-5em}
  \includegraphics[height=4cm]{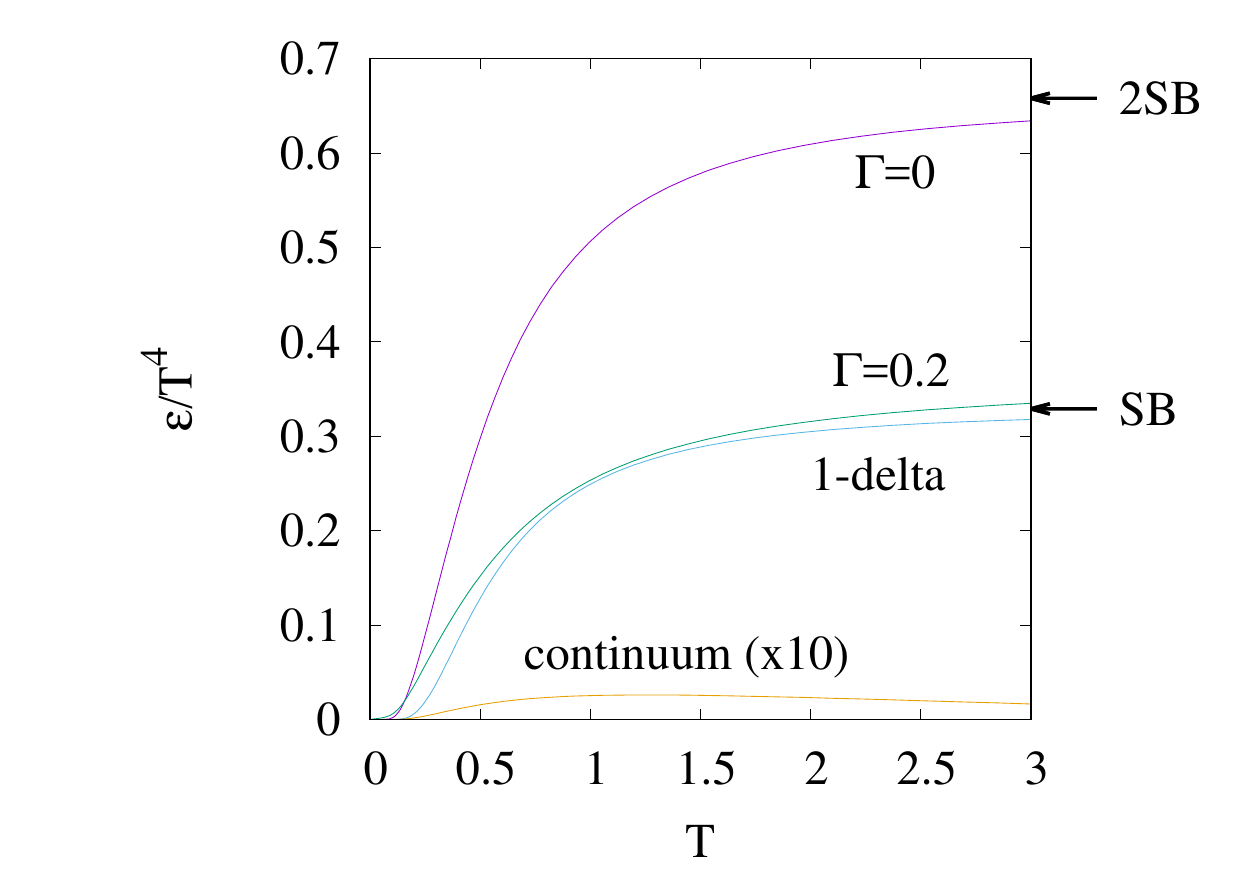}
  \caption{Left panel: spectral functions of two peaks with different
    widths as well as a pure continuum. Right panel: energy density of
    two peaks with different widths, the energy density of a single
    Dirac-delta peak (with $m=1.6$), and the energy density
    contribution of the continuum, enlarged by a factor of 10}
  \label{fig:merge}
\end{figure}

In the free case, ie. using the free spectral function
\eqref{eq:freespect} and ${\cal K} = p^2-m^2$ in the above formula, we
get back the free pressure \eqref{eq:freeP}. If there are several
Dirac-delta peaks in the spectrum, we also get back the sum of the
free gas partial pressures, in accordance with the Beth-Uhlenbeck
formula \eqref{eq:BethUhlenbeck}. But if the peaks are not
independent, the exact pressure starts to deviate from the
Beth-Uhlenbeck prediction. In Fig.~\ref{fig:merge} we can see that
when two peaks start to merge, the exact pressure decreases.
It is not unexpected: when there is just one peak, the
spectrum looks like a one-quasiparticle spectrum, and so the pressure
must come from a single degree of freedom. Therefore starting from a
two separate peaks spectrum, and continuously approach a one-peak
spectrum the pressure also changes smoothly from the two-particle
pressure to the one particle one.

This observation leads to the explanation of \emph{Gibbs-paradox} in
interacting systems \cite{Jakovac:2012tn}. The original paradox, valid in
free systems is that if the molecules of two gases differ only in a
tiny, continuously disappearing thing (eg. mass difference, or a tiny
``flag''), then the two gases are different as long as the difference
is present, but are the same if the difference is exactly zero. This
leads to a non-analytic contribution to the entropy (mixing
entropy). In interacting gases, however, the energy spectrum consists not
infinitely thin Dirac-deltas, but there is a line broadening coming
from different sources (eg. thermal motion, or finite density). Then
with vanishing mass difference the spectral functions become more and
more overlapping, as it is shown in Fig.~\ref{fig:merge}. As a
consequence the pressure will continuously reduce from the
2-independent particle pressure to a 1-particle pressure (where the
mass is somewhere between the masses of the two peaks), as it is
also shown in Fig.~\ref{fig:merge}. In an interacting system,
therefore, the Gibbs-paradox leads to a continuously vanishing mixing
entropy.

In Fig.~\ref{fig:merge} there is shown also the contribution of the
multiparticle continuum (cut) part. Its spectrum is not
quasiparticle-like, as it can be seen on the left panel of
Fig.~\ref{fig:merge}. The corresponding pressure is much lower than
the pressure of the quasiparticle systems: in Fig.~\ref{fig:merge} it
is enlarged by a factor of 10 to be visible at all.

This explains why a huge pressure reduction appears when a peak merges
with the continuum, ie. when it melts. The original narrow peak
structure corresponds to an almost free quasiparticle, with pressure
close to the free pressure. When the peak gets merged in the
continuum, the spectrum does not contain a particle, it becomes more
and more like the continuum part of Fig.~\ref{fig:merge}, therefore
the corresponding pressure is also smaller and smaller. In the course
of a continuous merging procedure the pressure smoothly changes from
the one free particle pressure to zero: the particle is melted, it
disappears from the thermal ensemble. \emph{We can define the number of
thermodynamical degree of freedom as the ratio of the exact pressure
and the free pressure}. With this definition the thermodynamical degrees
of freedom changes continuously to zero.

\begin{figure}[htbp]
  \centering
  \includegraphics[width=0.5\textwidth]{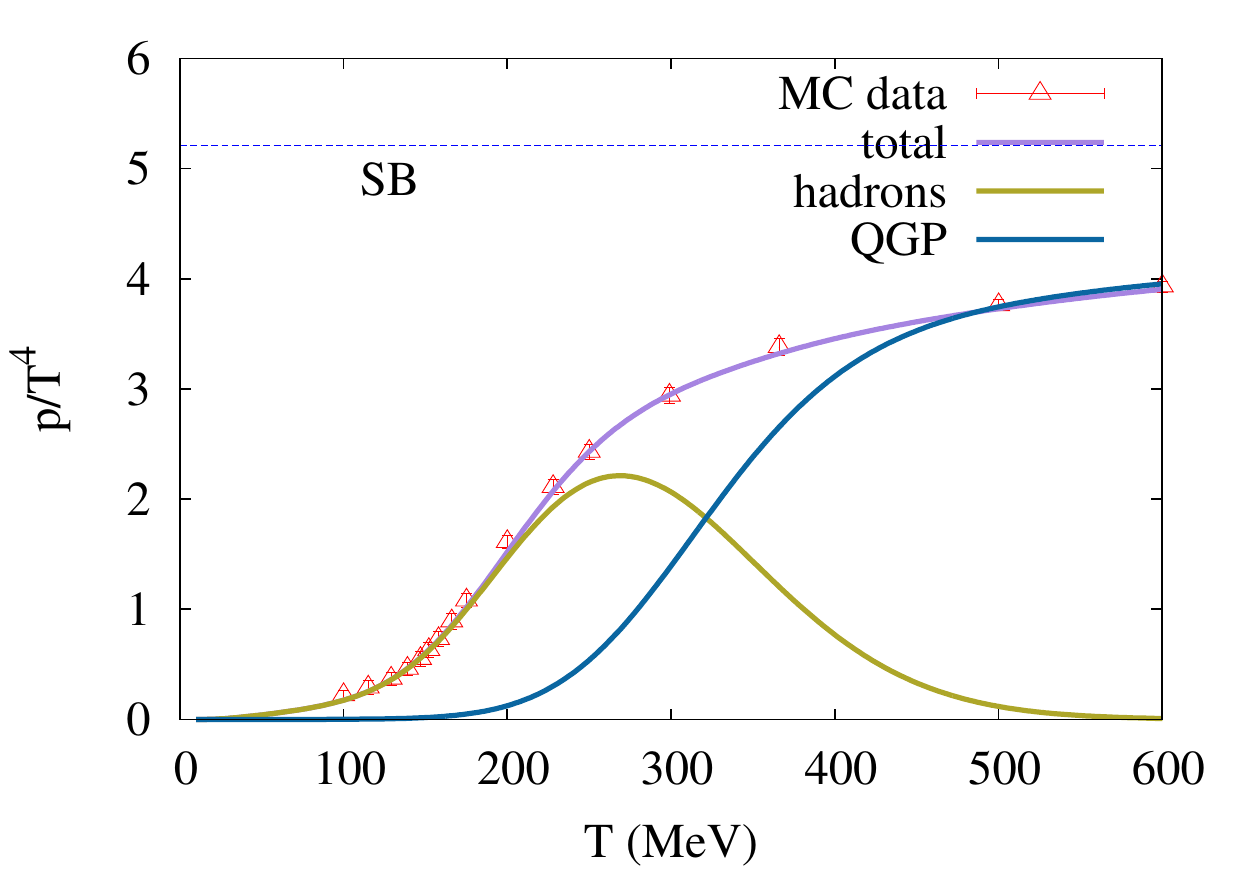} 
  \caption{Hadronic pressure reduction at high $T$, quark pressure reduction at low $T$,
	due to merging peaks. For comparison, lattice data are indicated by triangles.}
  \label{fig:QCDpressure}
\end{figure}

This mechanism makes it possible to explain why there is no divergent
pressure beyond the Hagedorn temperature in the QCD plasma, or why the
hadronic pressure does not overshoot the QGP pressure. The hadron
spectrum changes with the temperature from quasiparticle peaks to
peaks merged with the continuum. As discussed above, this results in
the reduction of the thermodynamical degrees of freedom effective for the
total pressure. The situation
is just the opposite for the QGP degrees of freedom: at small temperatures
the spectrum in the quark channel is just a continuum, there are no
particle-like excitations there, the partial pressure is zero. At high
temperatures the spectra in the QGP channels become more and more
particle-like; although even at about $T \approx 300$ MeV the number of
thermodynamical degrees of freedom is only about 80\% of the free case
(cf. Fig.\ref{fig:QCDpressure}).

In this way, within the picture of melting quasiparticle peaks, the QCD
pressure computed in MC simulations can be reproduced and interpreted
in correct physical terms. The main
prediction of this model is that the hadronic
thermodynamic degrees of freedom do not vanish suddenly above the
critical temperature, there is a sizable temperature regime, until about $T=330$ MeV,
where they still dominate the pressure. In this
melting hadron peak regime, however, we do not have quasiparticles as
excitations, just a mixture of hadron-like and dissociated
quark-gluon-like behavior.
This is indeed a new type of nuclear matter.

\subsection{Continuous mass fits to lattice EoS}



Quark matter, searched for in relativistic heavy ion collisions, reveals
itself in signatures on observed hadron spectra which are 
interpreted in terms of quark level properties. In particular scaling
of the elliptic flow component $v_2$ with the constituent quark content
of the finally observed mesons and baryons \cite{v2,v2B,v2C,v2D,v2E,v2F} and successful
description of $p_T$-spectra of pions and antiprotons using quark
coalescence rules for hadron building \cite{ourJPG} utilize the fast
hadronization concept of quark redistribution.
Albeit this simple idea brings also problems with it, e.g. in dealing
with energy conservation and entropy increase, these issues can be
resolved by using a distributed mass quasiparticle model for quark
matter \cite{ourPRChep}, and are in accord with the quark matter
equation of state obtained in lattice QCD calculations \cite{ourPLB}. 
The surmised mass distribution gives
rise to specific equation of state (pressure as a function of temperature, $p(T)$),
and reversed, a mass distribution may be outlined from knowledge on the
$p(T)$ curve.

While traditional, fixed mass quasiparticle
models already succeed to describe the equation of state obtained in
lattice QCD \cite{Quasiptl}, those mass values are themselves
temperature dependent. Furthermore a temperature dependent width
is associated to the quasiparticle mass, too \cite{Quasi2,Quasi3,Peshier}. 
The factor between the massive and massless relativistic ideal pressure
in the Boltzmann approximation,
\be
\Phi\left(\frac{m}{T}\right) \: = \: \frac{p_{{\rm id}}(m,T)}{p_{{\rm id}}(0,T)} 
 \: = \: \frac{1}{2} \, \frac{m^2}{T^2} \, K_2\!\left(\frac{m}{T}\right),
\ee{PRESRATIO}
relates the observed pressure in a non-trivially interacting system to the
mass distribution of a conjectured continuous mixture of different mass particles
\be
 p(T) \: = \: \infi\! w(m) \, p_{{\rm id}}(m,T) \, dm.
\ee{TOTALPRESS}
This means that the observed equation of state in terms of the pressure ratio to the Stefan--Boltzmann
limit, representing the effective fraction of thermodynamical degrees of freedom,  
is a so called Meijer transform of a conjectured continuous mass distribution:
\be
\sigma(T) \: = \: \frac{p(T)}{p_{{\rm id}}(0,T)} 
 \: = \: \infi\!\! w(m) \, \frac{1}{2} \,  \frac{m^2}{T^2} \, K_2\!\left(\frac{m}{T}\right) \, dm. 
\ee{PRESSASMEIJER}
Using the scaled variables, $t=m/T_c$ and $z=T_c/T$, and the redefined functions
$\sigma(z):=\sigma(T_c/z)$, $w(t):=T_cw(T_ct)$, we obtain
\be
\sigma(z) \: = \: \infi\!\! w(t) \, \frac{z^2t^2}{2} \, K_2\!\left(zt\right) \, dt. 
\ee{SCALEDPRESSASMEIJER}
This integral transformation, the so called Meijer transform, can be inverted analytically,
\be
 w(t) \: = \: \frac{2}{i\pi} \int \sigma(z) \, \frac{I_2(zt)}{zt} \,  dz.
\ee{MASSDISTINVMEIJER}
The respective high temperature expansions of the pressure ratios, based on the expansion
of the $K_2(z)$ Bessel function in the mass distribution formula,
and that one applied in perturbative QCD, are worth to be compared: 
\ba
 \frac{p(T)}{p_{{\rm id}}(0,T)} & = & 
 1 - \frac{\exvs{m^2}}{4T^2}+ \left(\frac{3}{4}-\gamma\right)\frac{\exvs{m^4}}{16T^4}
 \nl
 & & + \, \frac{\exvs{m^4\ln(2T/m)}}{16T^4} +\ldots, \nl
 \nl
 \frac{p_{{\rm pQCD}}}{p_{{\rm id}}(0,T)} &=& 1 - a_2g^2 +a_4g^4 +b_4g^4\ln\frac{2\pi T}{\Lambda}+\ldots 
\ea{pQCD}
with $\gamma$ being the Euler--Mascheroni constant and $\Lambda$ the renormalization subtraction scale.
We note that $\sigma(z)$ can also be obtained for Bose or Fermi
distributions instead of the Boltzmann one; the numerical difference
is overall minor, less than six per cent at vanishing chemical potential. 
The basic result on the Debye screening length in a QGP supports the
assumption that $\exv{m^2}\sim g^2T^2$
sets the scale for a simplified treatment of the quark matter pressure at high
temperature.
The comparison of pQCD and mass distribution results above
reveals that $\exv{m^4} \ne \exv{m^2}^2$, whence the necessity of
a width in the mass distribution emerges. 
{\em Alone this fact indicates that the spectral function cannot be a simple sum of quasiparticle peaks,
it must contain appreciable widths, possibly even a continuum part.}

The temperature dependence of the
pressure ratio to the massless ideal gas value is concentrated on the
temperature dependence of the coupling constant: $g=g(T)$ in the traditional
interpretation. We have recently pursued an alternative approach to the
quasiparticle mass distribution in quark matter \cite{ourJPG,ourPRChep}, where
a temperature independent $w(m)$ distribution is reconstructed from the
pressure ratio $\sigma(T)=p(T)/p_{{\rm id}}(0,T)$ curve:
\be
 \sigma(T) = \int_0^{\infty} w(m) \: \Phi\left(\frac{m}{T}\right) \, \d m.
\ee{PRESS}
It is interesting to play around with some analytic formulas with respect to the Meijer
transformation. The simple exponential ansatz leads to a certain power-law tailed
form of $w(m)$ with a threshold mass gap at $m= \lambda$:
\be
 \sigma(T) \: = \: \ead{-\lambda/T} \: \leftrightarrow \:
 w(m) \: = \: \frac{4\lambda}{\pi m^2} \, \sqrt{1-\frac{\lambda^2}{m^2}}.
\ee{SIGMAW_EXP}
Since such a mass distribution would have a diverging $\exv{m^2}$ and also for
higher powers of $m$, we conclude that it must be
\be
 \sigma(T) \: < \: \ead{-\lambda/T}.
\ee{LESSTHANSIGMA}
Indeed lattice results on $\sigma(T)$ all satisfy such a constraint with a corresponding
value of $\lambda$. The smallest such $\lambda$ value, found numerically, is then the
Boltzmannian estimate for the mass gap.
It is a remarkable property of this approach
that it indicates a temperature independent threshold (smallest mass) in the
$w(m)$ spectrum for lattice QCD pressure data\cite{FODOR,BIELEFELD}.

The pressure is, however, not known analytically, the numerical results are
smeared with error bars. This problem is more severe in the light of the
fact that eq.(\ref{PRESSASMEIJER}) constitutes an integral transformation 
(the Meijer K transformation, a generalization of the Laplace transformation).
There is no mathematical guarantee that the inverting  transformation eq.(\ref{MASSDISTINVMEIJER})
leads to close results for $w(m)$ from close functions for $\sigma(T)$.
In fact this is known as the ''inverse imaging problem'' \cite{IMAG,IMAG2,IMAG3}.

\begin{figure}[htb]

\includegraphics[width=0.28\textwidth,angle=-90]{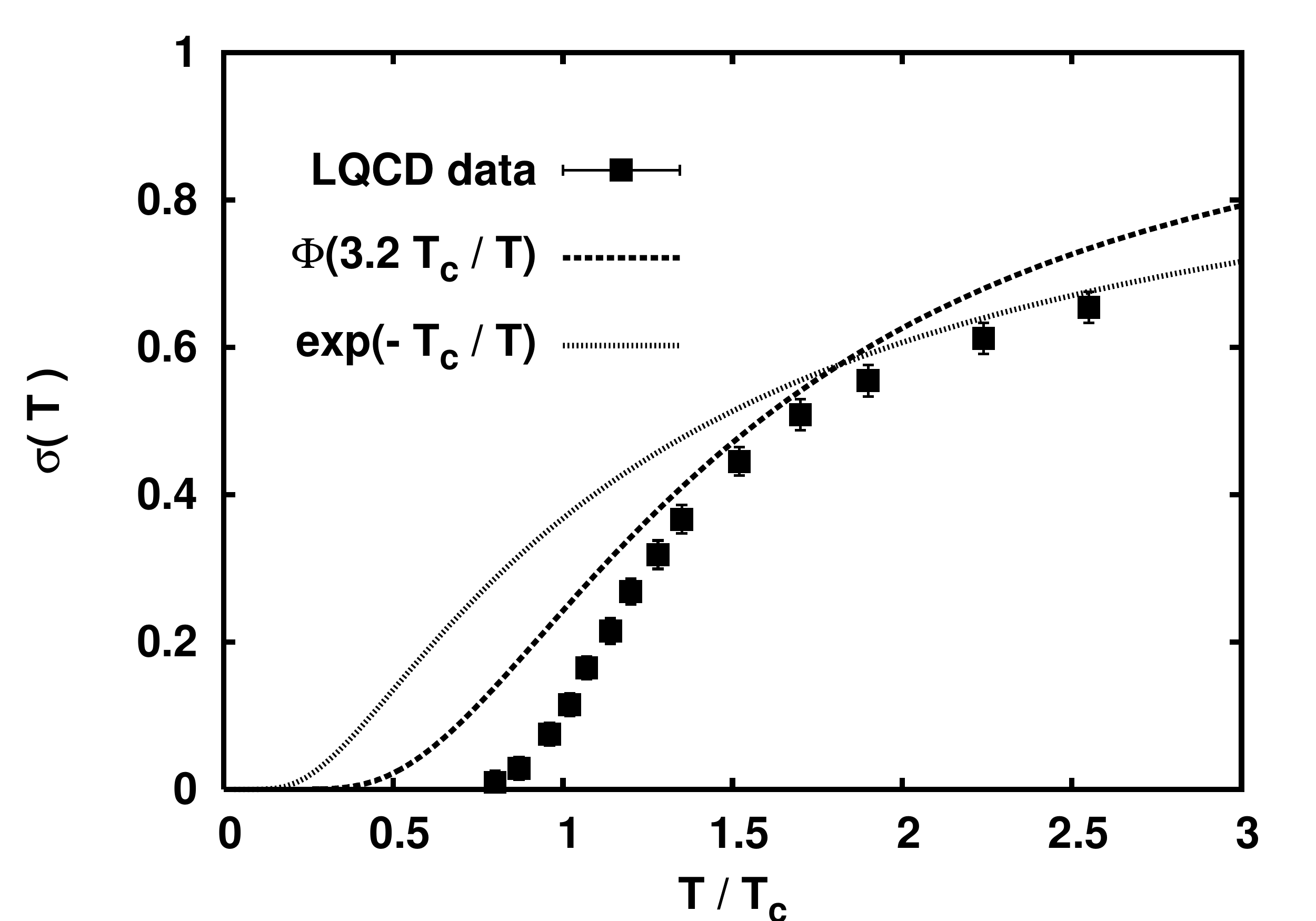}
\includegraphics[width=0.28\textwidth,angle=-90]{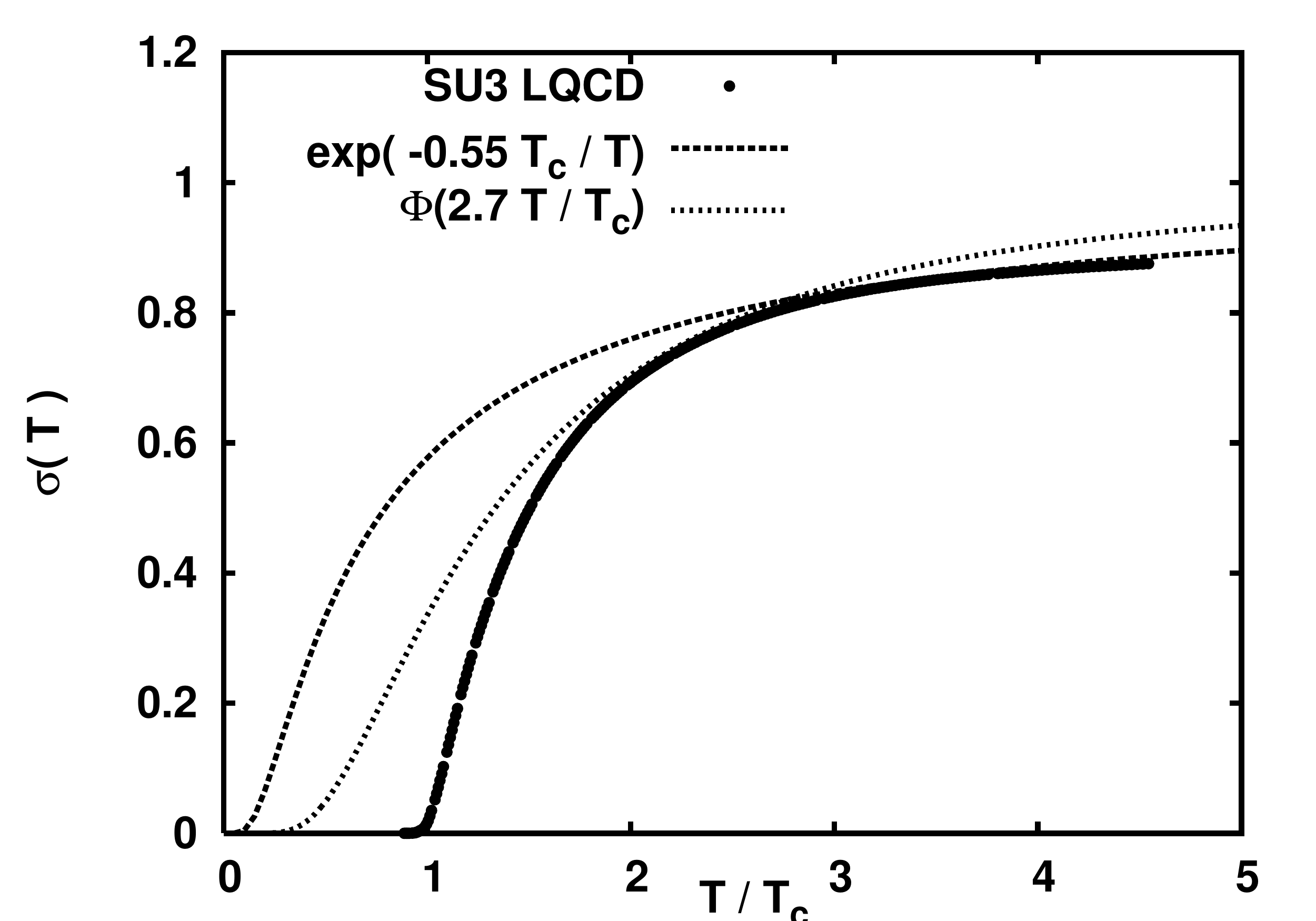}

\caption{ \label{FIG0}
  Pressure curve estimators for the data from lattice QCD simulation of Ref.\cite{FODOR}
  (2+1 flavor QCD) and of \cite{SU3,SU3B} (pure SU(3) gauge theory).
}
\end{figure}

However, based on the above assumptions
one can obtain some supportive knowledge about a $T$-independent $w(m)$
mass distribution when the pressure $p(T)$ satisfies certain inequalities.
In particular we prove that if the pressure $p(T)$ is below the corresponding
ideal gas pressure with a given mass $M_0$ at {\em all temperatures}, then the
mass distribution is exactly zero for all masses below $M_0$. 
For inequalities with other than ideal gas pressure curves as
estimators we apply the Markov inequality for probability measures,
which directly offers upper bounds on the integrated probability density function
$P(M)=\int_0^M\limits w(m) dm$.  
It turns out that the appearance and value of the highest possible
$M$ for which $P(M)=0$, the mass gap value $M_0$, 
is connected to the low temperature behavior of $\sigma(T)$.
Two particular estimators for $\sigma(T)$, namely $\Phi(M_0/T)$ with $M_0=3.2 T_c$ and
$\exp(-T_c/T)$ are compared to 2+1 flavor lattice QCD scaled pressure data
in Figure \ref{FIG0} (top) and to pure SU(3) lattice gauge theory data (bottom) with
$M_0=2.7T_c$ and $\lambda=0.55T_c$.
Of course the temperature scales are different, $T_c\approx 165$ MeV in the first,
$T_c \approx 260$ MeV in the second case. 
These examples are important for gaining a physical insight
into the Markov inequality discussed below.

In the followings we relate the mass gap to the behavior of $\sigma(T)$
using the generalized Markov inequality
to estimate upper bounds on the integrated probability density function
for the mass being lower than a given value.
The general form of the Markov inequality is given by \cite{MARKOV1,MARKOV2,MARKOV3,MARKOV4}
\be
 \mu\left(\left\{x\in X: f(x) \ge t \right\} \right) \le \frac{1}{g(t)} 
  \int\limits_{x\in X} g\left(\,f(x)\,\right) \d\mu(x)
\ee{MARKOV}
with measure $\mu$, a real valued $\mu$-measurable function $f$,
and a monotonic growing non-negative measurable real function $g$. 
The proof, based on the monotonity of integration, 
can be presented in a few lines. For a non-negative and monotonic growing
function $g(t) \le g(f(x))$ for $t \le f(x)$. We obtain
\be
 g(t)\!\!\!\!\!\!\int\limits_{f(x)\ge t}\!\!\!\!\!\!\d\mu(x)  \: =  
 \!\!\!\!\!\!\int\limits_{f(x)\ge t}\!\!\!\!g(t) \d\mu(x) \: \le 
 \!\!\!\!\!\!\int\limits_{f(x)\ge t}\!\!\!\!g\left(\, f(x)\, \right) \d\mu(x).
\ee{MPROOF1}
This quantity can be bounded by 
\be
 \int\limits_{f(x)\ge t} g\left(\, f(x)\, \right) \d\mu(x) \: \le \:
 \int\limits_{x\in X} g\left(\, f(x)\, \right) \d\mu(x).
\ee{MPROOF2}
A division by $g(t)\ge 0$ delivers the original statement in eq.(\ref{MARKOV}).

In order to apply this inequality to the mass spectrum we choose $f(m)=tM/m$.
In this case
\be
 \mu \left(\frac{tM}{m} \ge t \right) = \int_0^{M} \d\mu(m),
\ee{WHATmu}
and the Markov inequality reads as
\be
 P(M) := \int_0^{M} \d\mu(m) \le \frac{1}{g(t)} \int_0^{\infty}g\left(\frac{tM}{m}\right) \d\mu(m).
\ee{OTHER-MARKOV}
For a continuum mass spectrum $\d\mu(m)=w(m)\d m$ can be chosen with $w(m)$ being the
probability density function. The generalized Markov inequality stated above is
valid for general probability measures\footnote{a measure normalized to one} 
$\mu$ possibly including bound state contributions.


Now we discuss a few examples for monotonic rising functions $g(z)$, which
allow us to draw some conclusions about the integrated probability for masses below
$M$.  Applying the special form of $g(t)=t^n$ we arrive at
\be
 P(M) \le \frac{1}{t^n} \int_0^{\infty}\left(\frac{tM}{m}\right)^n w(m) \d m,
\ee{PSMALLER}
whence we obtain:
\be
 P(M) \le M^n \int_0^{\infty} m^{-n} w(m) \d m.
\ee{ourMARKOV}
It is easy to see that the negative integral moments of the mass on the right hand side of
the above inequality are connected to the negative integral moments of scaled pressure
$\sigma(T)=p(T)/p_{{\rm id}}(0,T)$. The final inequality for the probability of having
masses smaller than $M$ is given by
\be
 P(M) \: \le \: M^n \, \frac{\infi T^{-n-1} \, \sigma(T) \, \d T}{\infi x^{n-1} \, \Phi(x) \, \d x}.
\ee{AHA}

\begin{figure}[thb]

\includegraphics[width=0.28\textwidth,angle=-90]{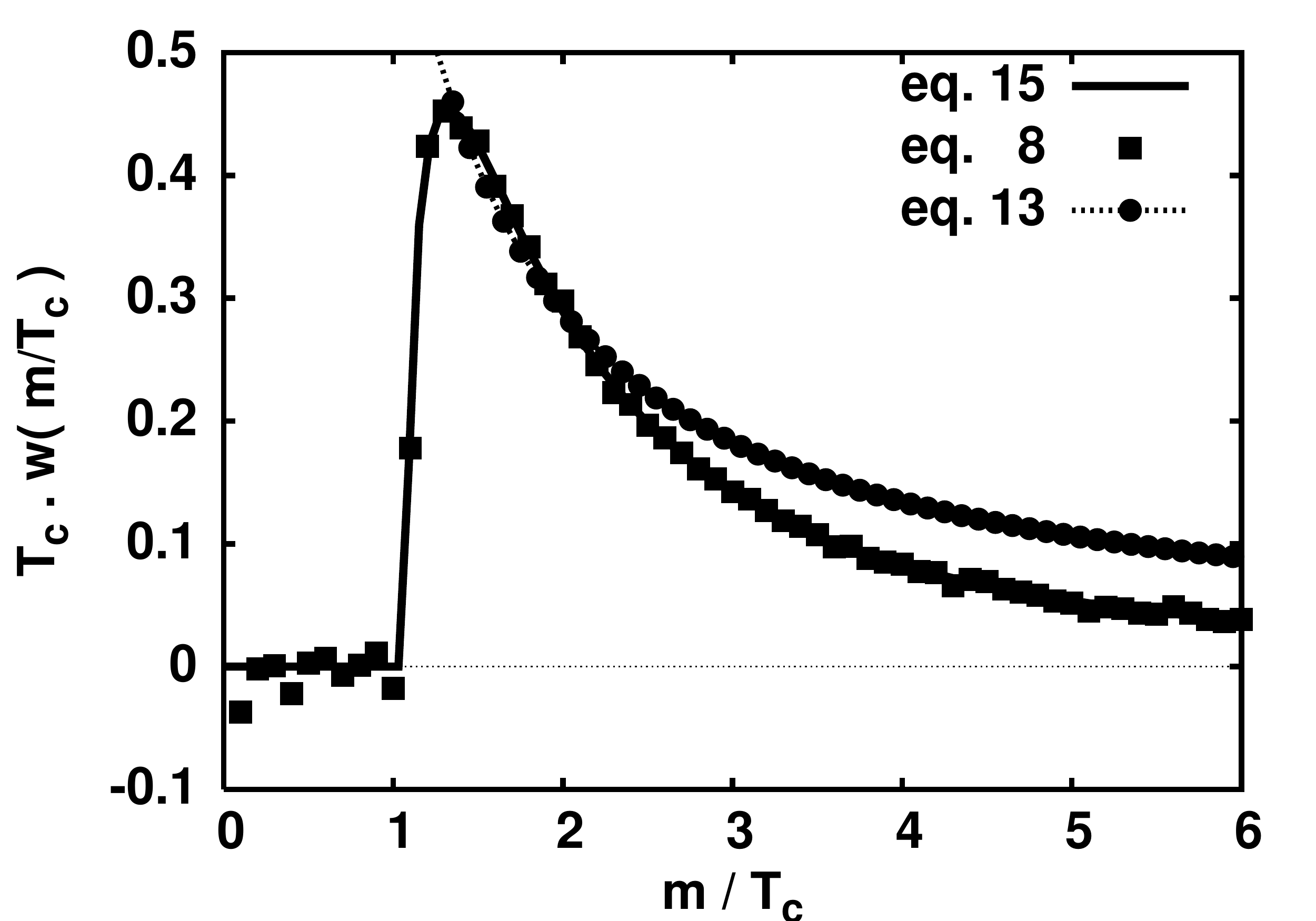}
\includegraphics[width=0.28\textwidth,angle=-90]{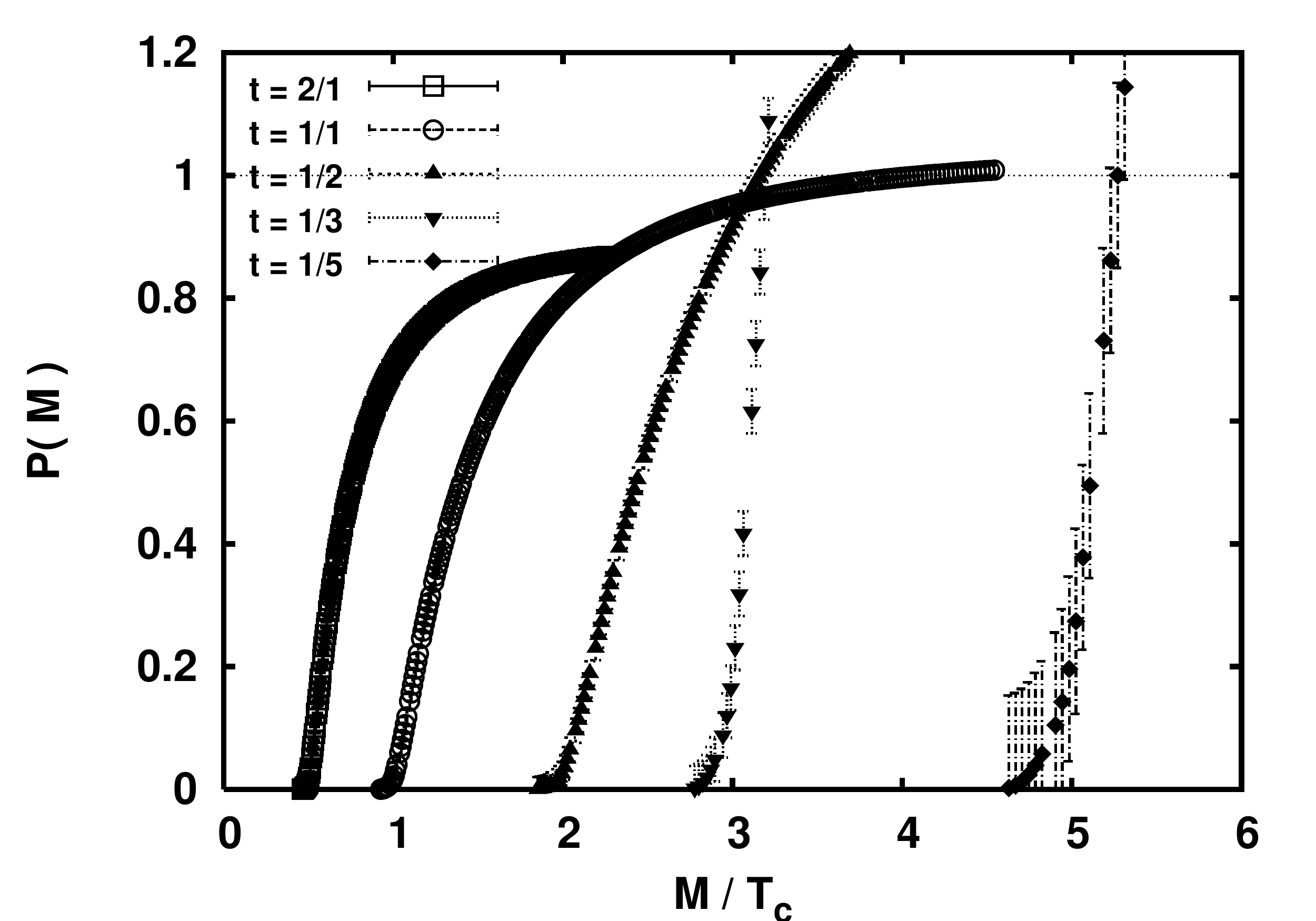}

\caption{ \label{FIG-PM}
Reconstructed scaled mass distribution using analytic (continuous line) and numerical
Meijer back transformation (black boxes: lattice EoS data, black circles: exponential upper estimate) 
on the top.
Upper bounds for the integrated probability $P(M)$ of masses lower than $M$,
based on 2+1 flavor lattice QCD EoS data\cite{FODOR} at the bottom 
(cf. eq.\ref{STRIKING}).
}
\end{figure}

Let us apply this result to the simplest majorant, that of a fixed mass relativistic ideal gas.
In this case $\sigma(T)\le \Phi(M_0/T)$ with some $M_0$ (cf. dashed line in Figure \ref{FIG0}). 
Equation (\ref{AHA}) leads to
\be
 P(M)\: \le \: M^n
 \frac{\int_0^{\infty}T^{-n-1} \, \Phi\!\left(\frac{M_0}{T}\right) \d T }{\int_0^{\infty}x^{n-1}\Phi(x) \d x}
 \: \le \: \left(\frac{M}{M_0}\right)^n
\ee{IDEAL}
in this case. Should it hold for arbitrary high $n$, the right hand side of this inequality
is zero for all $M<M_0$ and divergent for $M>M_0$. In the second case it is not restrictive,
since $P(M)<1$ anyway, in the first case this means a mass gap up to $M_0$.
We note that this conclusion holds for a general non-negative $\Phi(x)$, for which the 
integrals 
in eq.(\ref{IDEAL}) are finite for all $n>0$. Thus the Bose-Einstein or the Fermi-Dirac
distribution could as well be applied instead of the Boltzmann one.


Another possible majorant is the exponential function, $\sigma(T) \le \exp(-\lambda/T)$
(cf. the dotted line in Figure \ref{FIG0} for $\lambda=T_c$). 
In this case eq.(\ref{AHA}) delivers
\be
 P(M) \le \left(\frac{M}{2\lambda}\right)^n \frac{2\Gamma(n)}{\Gamma(2+n/2)\Gamma(n/2)}.
\ee{EXP}
The large $n$ limit of this result is given by
\be
 P(M) \le \left(\frac{M}{\lambda}\right)^n 2\sqrt{\frac{2}{\pi}} 
  \frac{1}{n^{3/2}}
\ee{LARGE-N-EXP}
to leading order in $1/n$. Again the right hand side approaches zero for $M\le \lambda$
and diverges for $M > \lambda$. This points out a mass gap stretching to 
(and including) $\lambda$ from zero.

\begin{figure}[hb]

\includegraphics[width=0.28\textwidth,angle=-90]{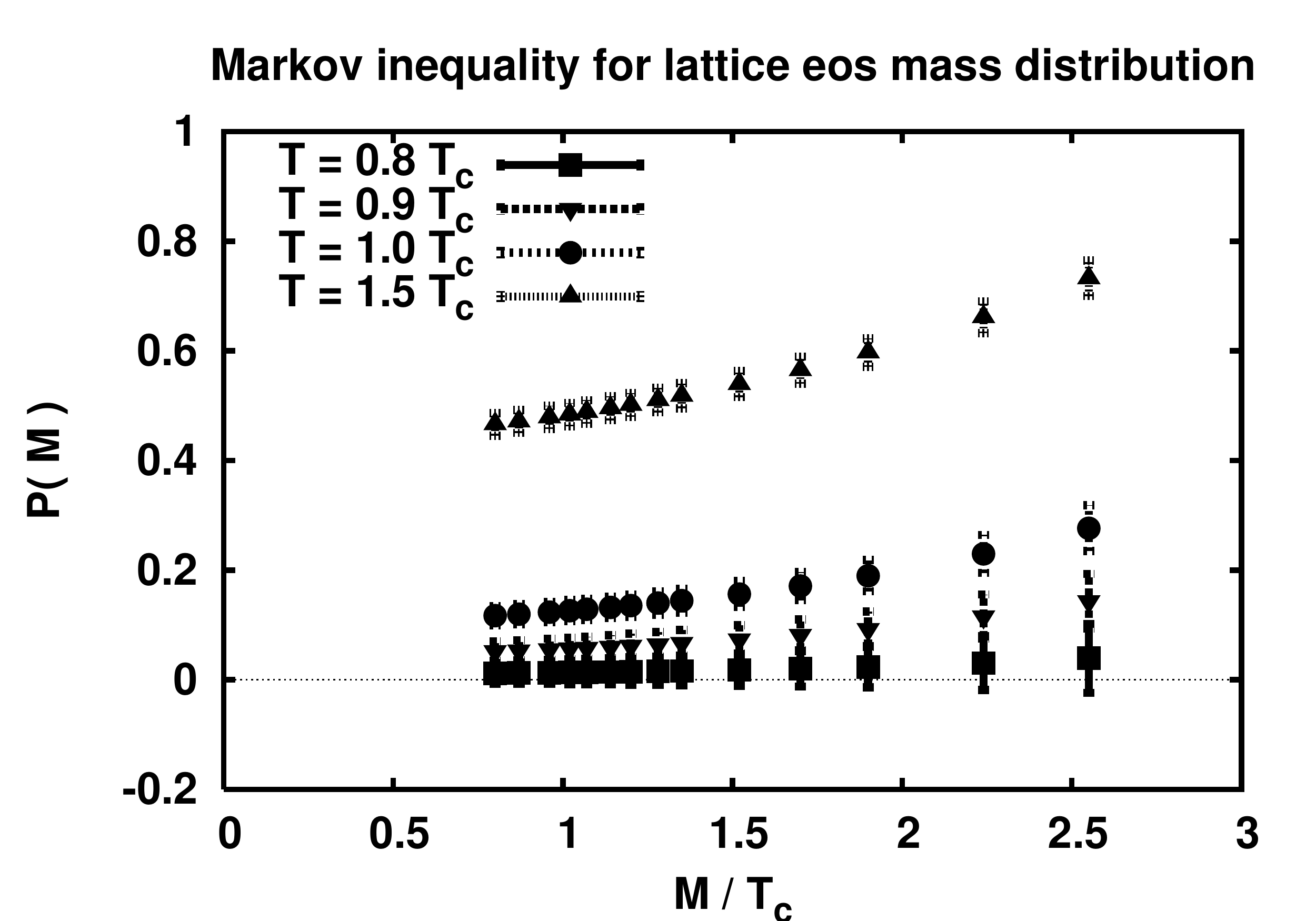}
\includegraphics[width=0.28\textwidth,angle=-90]{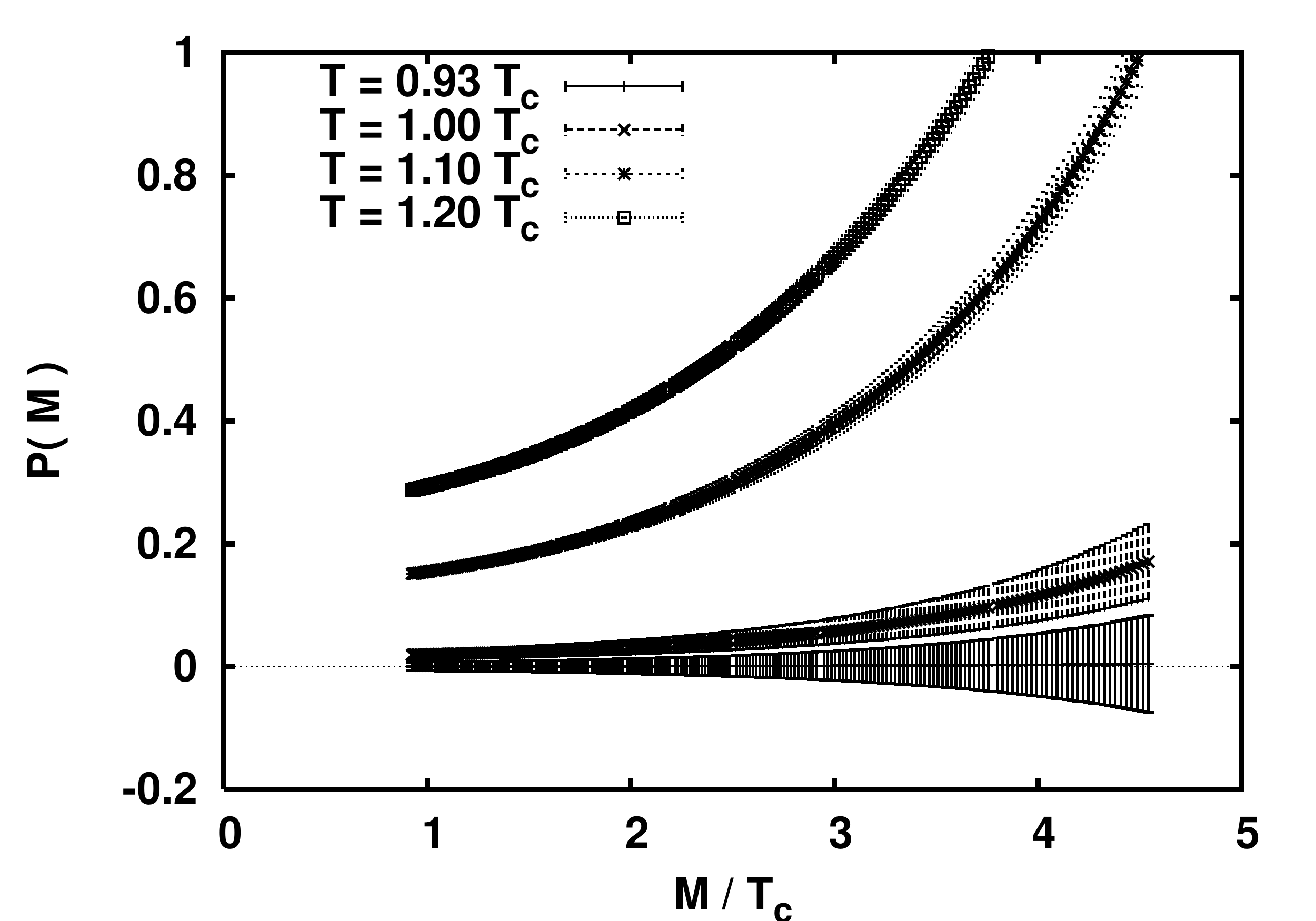}

\caption{ \label{M2}
Upper bounds for the integrated probability of masses lower than $M$ based
on eq.(\ref{PRACTICAL}) and
on 2+1 flavor lattice QCD EoS data\cite{FODOR}(top) and from Ref.\cite{BIELEFELD} (bottom)
respectively.  
The estimates belong to different temperatures  $T$ near $T_c$
indicated in the legend. In the second case a constant error of $0.01$ was assumed
in the original $p/T^4$ data.
}
\end{figure}


The most striking inequality is obtained by using $g(t)=\Phi(1/t)$.
This function is also admissible, its rise from zero to one is strict monotonic.
Eq. (\ref{AHA}) leads to
\be
 P(M) \: \le \: \frac{\sigma(tM)}{\Phi(1/t)}.
\ee{STRIKING}
For $t=1$ using the numerical value $\Phi(1)\approx 0.81$ one arrives at
$P(M) \le 1.23 \sigma(M)$, which can be directly read off from numerical
simulation or theoretical predictions of $\sigma(T)$.
Figure \ref{FIG-PM} presents curves for different $t$-values (see legend), all being
an upper estimate for the integrated probability $P(M)$ in the respective cases of
2+1 flavor QCD and pure SU(3) gauge theory. The higher seems to be the starting
$M_0$ value for the rise of the upper bound on $P(M)$, the higher also
the magnification of the error bars. A secure estimate for the $P(M)\le 0.05$
is given for masses $M > 1.7 T_c = 280$ MeV for the 2+1 flavor QCD case,
while for $M > 7.2T_c = 1.9$ GeV for the pure SU(3) gauge case. While in the first case
this can be at best an average between quark and gluon-like quasiparticle masses,
in the second case should be close to observed glueball mass.
We note that using $\sigma(tM) \le \Phi(M_0/tM)$ in the $t \rightarrow 0$ limit
again a mass gap at $M_0$ follows from eq.(\ref{STRIKING}). In this respect
the use of different $g(z)$ functions in the Markov inequality does not 
matter\footnote{From practical viewpoint, however, in the $t\rightarrow 0$ limit
the error bars on the original $p/T^4$ data are infinitely enlarged.}.

A related version of the inequality (\ref{STRIKING}) is obtained for $tM=T$,
$g(z)=\Phi(tM/zT)$.
The upper bound is obtained at any fixed $T$ as being
\be
 P(M) \: \le \: \frac{\sigma(T)}{\Phi(M/T)}
\ee{PRACTICAL}
Figure \ref{M2} plots upper bounds for $P(M)$ obtained using the
eq.(\ref{PRACTICAL}). The most restrictive are the lowest temperature
data for $\sigma(T)$, they are, however, also the most contaminated
by errors. It is probably safe to conclude that as much as
$90-95\%$ of the masses are above $1.5 T_c \approx 440$ MeV according to
these data.

Our mathematical treatment of the mass gap leaves the point $m=0$ in the
possible mass distribution as a special case. Assuming that there were
such a contribution of finite measure, i.e. $P(0)=a$ were a finite
value between zero and one, one concludes from the definition
eq.(\ref{PRESS}) that in this case $\sigma(T) \ge a$ would be. There is no
sign of such an indication in lattice QCD data.

Finally we note that there is a potential to use our method presented in this
letter in a context wider than quark matter: the quasiparticle test
based on the generalized Markov inequality can in principle be done for
any system with sufficiently known thermal equation of state. The estimate
for a lowest mass can then be checked against knowledge on the mass
spectrum obtained from the study of correlation functions.


\subsection{$\langle \alpha(Q^2) \rangle_T$ vs $\alpha_{{\rm eff}}(T)$}

%

Talking about non-perturbative effects in high temperature QCD, at a first glance is a paradoxical
issue. However, there always have been warnings coming from a few 
experts \cite{Polonyi1,Polonyi2,Polonyi3,Polonyi4}.
By the majority such warnings have been long ignored: upon the famous
proof by A.~Linde \cite{LINDE}, that the problem of non-perturbativness were an infrared effect,
it was generally believed that one does not have to consider this above $T_c$.

The $1/log$-like pole behavior of the running coupling constant has been encountered
by phenomenological shifts in the renormalization point energy scale from $T_c$ a bit
\cite{SHIFTLOG1,SHIFTLOG2,SHIFTLOG3}   
in a formula for the effective thermal coupling
conjectured to be a good approximation:
\be
 \alpha_T :=  \exv{\alpha(Q^2)}_{{\rm thermal}} \: \approx \: 
 \alpha\left( \exv{Q^2}\right) = \alpha(T^2).
\ee{RUNALFA}
Even without digging into the delicate issues of QCD deep, one can easily convince 
himself that this approximation could only hold if the thermal distribution
of relative $Q^2$ values in a QGP were sharp. This is, however, not the case,
as we shall demonstrate it below.

First we summarize the results we arrive at by
considering the thermal distribution of $Q^2$ in a QGP:
\begin{enumerate}

\item	The thermal distribution of $Q^2$ values are not peaked
	around $T^2$, rather they are maximal at $Q^2=0$ between two massless
	particles; the Boltzmann distribution being just a particular example. 
	The width of the distribution is proportional to $T^2$.

\item	The expectation value of a non-perturbative (NP) order parameter,
	being one until $Q^2=\Lambda^2$ and zero otherwise,
	is non-vanishing at arbitrary temperatures. For high $T$ it
	goes like $\Lambda^2/T^2$ upon the constant probability near to
	$Q^2=0$.

\item	As a consequence at arbitrary high temperatures there is a
	relative measure of NP effects. In the pressure this occurs
	already at the subleading term.

\item	The lattice EoS results subleading terms are seen in the scaling
	$(e-3p)/T^2 =$ constant at high temperatures. pQCD would predict
	an inverse logarithmic fall of this value.

\end{enumerate}

In the followings we outline the support for these statements.
The relativistic kinematics for pairs of massless particles delivers
\be
 Q^2 =  - (p_1-p_2)^2 = 2E_1E_2(1-\cos\theta).
\ee{KINEM}
The distribution of $Q^2$ values is then given by
\be
 P(Q^2) = \frac{\int_{12}\limits \, f(E_1) f(E_2) \, \delta\!\left(Q^2-2E_1E_2(1-\cos\theta)\right)}{\int_{12}\limits \, f(E_1) f(E_2)}.
\ee{THERMDIST}
with the integration
\be
 \int_{12}\limits \ldots \: = \:  \int dE_1 dE_2 d\cos\theta \: E_1^2 \, E_2^2 \, \ldots
\ee{INTEGRAL12}
Using the Dirac delta functional for the integral over $\cos\theta$ this can be written as
\be
 P(Q^2) = \frac{\int_0^{\infty}\limits \!dE_1 \int_{Q^2/4E_1}^{\infty}\limits \!dE_2  \, \frac{1}{2}E_1E_2 f(E_1) f(E_2) }{\int_0^{\infty}\limits \!dE_1 \int_0^{\infty}\limits \!dE_2 \, 2E_1^2E_2^2 f(E_1) f(E_2)}.
\ee{PQ2}
This value is always between zero and one, its integral is one due to its construction  in
eq.(\ref{THERMDIST}). Since the thermal parton distribution, $f(E)$, is positive, the
numerator is maximal at $Q^2=0$. The maximal value of the distribution is given by
\be
 P(0) = \frac{1}{4} \frac{\int dE_1 \int dE_2 E_1E_2f_1f_2}{\int dE_1 dE_2 E_1^2E_2^2 f_1 f_2}
  =  \exv{\frac{1}{2E}}^2 = \frac{c^2}{T^2}
\ee{PQMAX}
with $c$ some constant depending on the distribution.
In particular for the Boltzmann distribution, $f(E)\propto e^{-E/T}$, the $Q^2$-distribution 
can be given in analytic form as
\be
 P(Q^2) =  \frac{1}{64T^2} \left(\frac{|Q|^3}{T^3} K_1(|Q|/T)+2\frac{Q^2}{T^2}K_2(|Q|/T) \right).
\ee{PQANAL}

\begin{figure}
\includegraphics[width=0.6\linewidth,angle=-90]{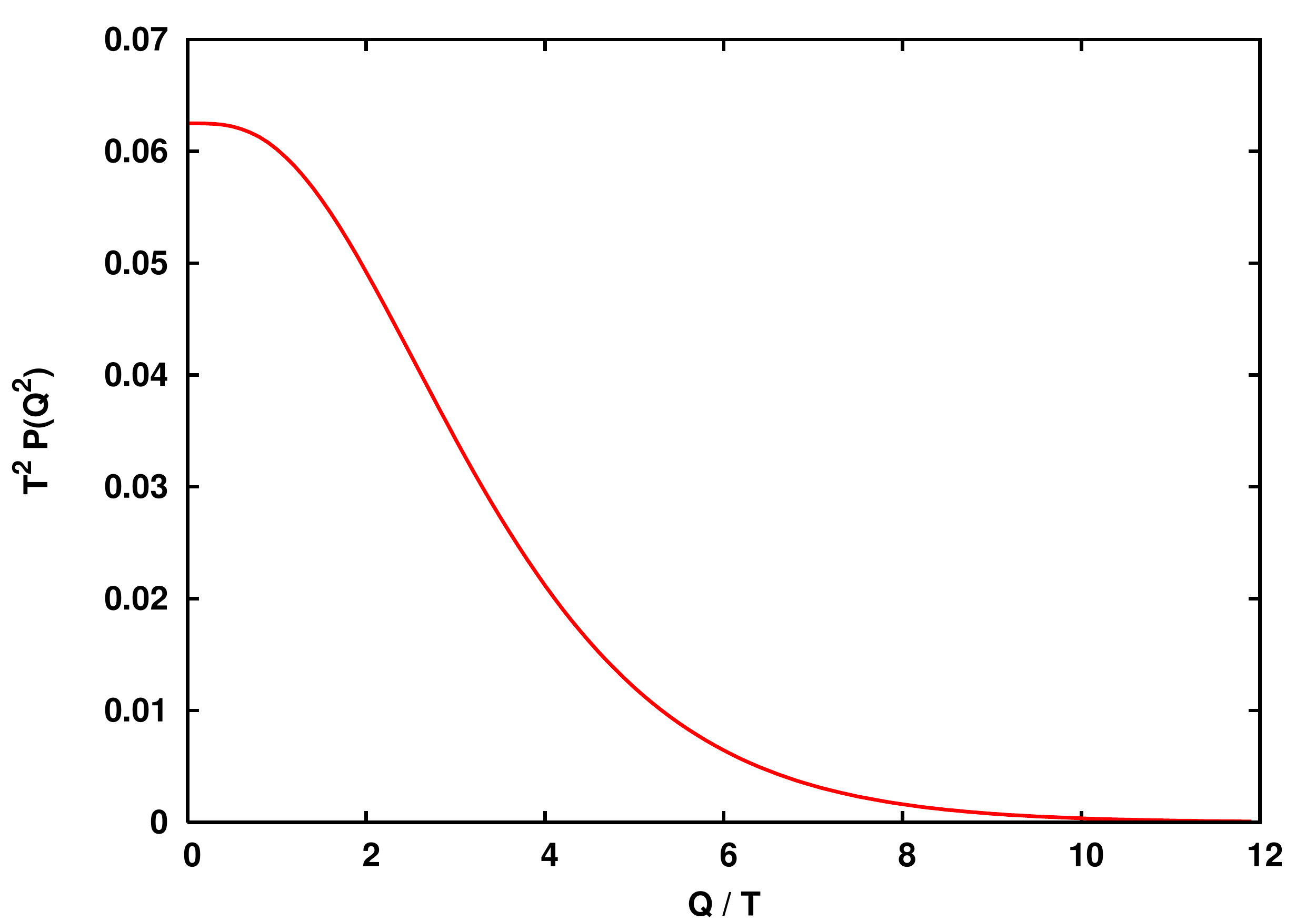}
\caption{\label{FIG1:PQ2}
 The distribution of $Q^2$ values between pairs of massless partons at temperature
 $T$ assuming respective Boltzmann distributions. It is not peaked, but its width scales with
 $T^2$.
}
\end{figure}
\vs
Let us now consider a non-perturbative quantity, like the string tension, 
which is zero above $Q^2=\Lambda^2$ and around constant below this momentum scale.
The ideal order parameter is given by a step function, $o = \Theta(\Lambda^2-Q^2)$.
The expectation value of such an order parameter at temperature $T$ is 
\be
 \exvs{o} = \int_0^{\Lambda^2}\limits\! d{Q^2} \, P(Q^2) = \int_0^{\Lambda^2/T^2}\limits\! F(x) dx
\ee{ORDER}
by using the integration variable $x=Q^2/T^2$. This is the integrated distribution
function of the thermal $Q^2$ distribution. By definition this approaches the value
one from below, so as a function of $T$ (or $T^2$) it starts with the value $1$ and
continuously decreases.

\begin{figure}
\hspace*{15mm}\includegraphics[width=0.6\linewidth,angle=-90]{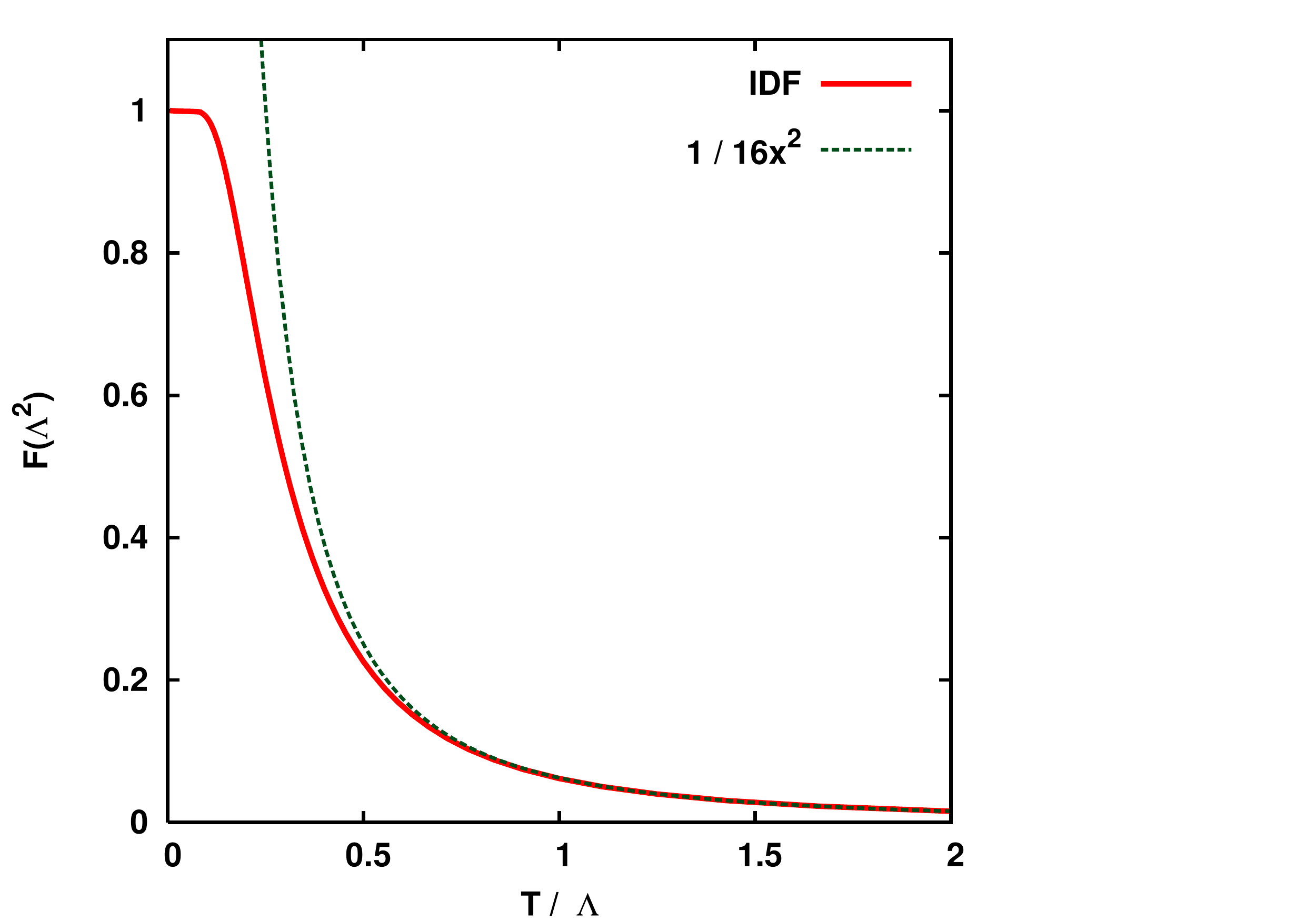}
\caption{\label{FIG2:InvIPQ2}
The integrated distribution of $Q^2$ values from zero to $\Lambda^2$
as a function of $T/\Lambda$ in order to show the contribution of an
NP order parameter. The dotted line is the high-temperature limit for
the Boltzmann distribution.
}
\end{figure}

For high enough temperature, $T \gg  \Lambda$, the distribution is nearly constant,
so we can use \eqref{PQMAX} to conclude that
\be
 \exvs{o} \approx \frac{c^2\Lambda^2}{T^2}.
\ee{HIGHT}
For Boltzmann distribution $c=1/4$.
This means that at any temperature there are non-perturbative (NP) effects
to subleading order in $T^2$ (cf. the $1/16x^2$ curve on Fig.\ref{FIG2:InvIPQ2}).

\begin{figure}
\includegraphics[width=0.6\linewidth,angle=-90]{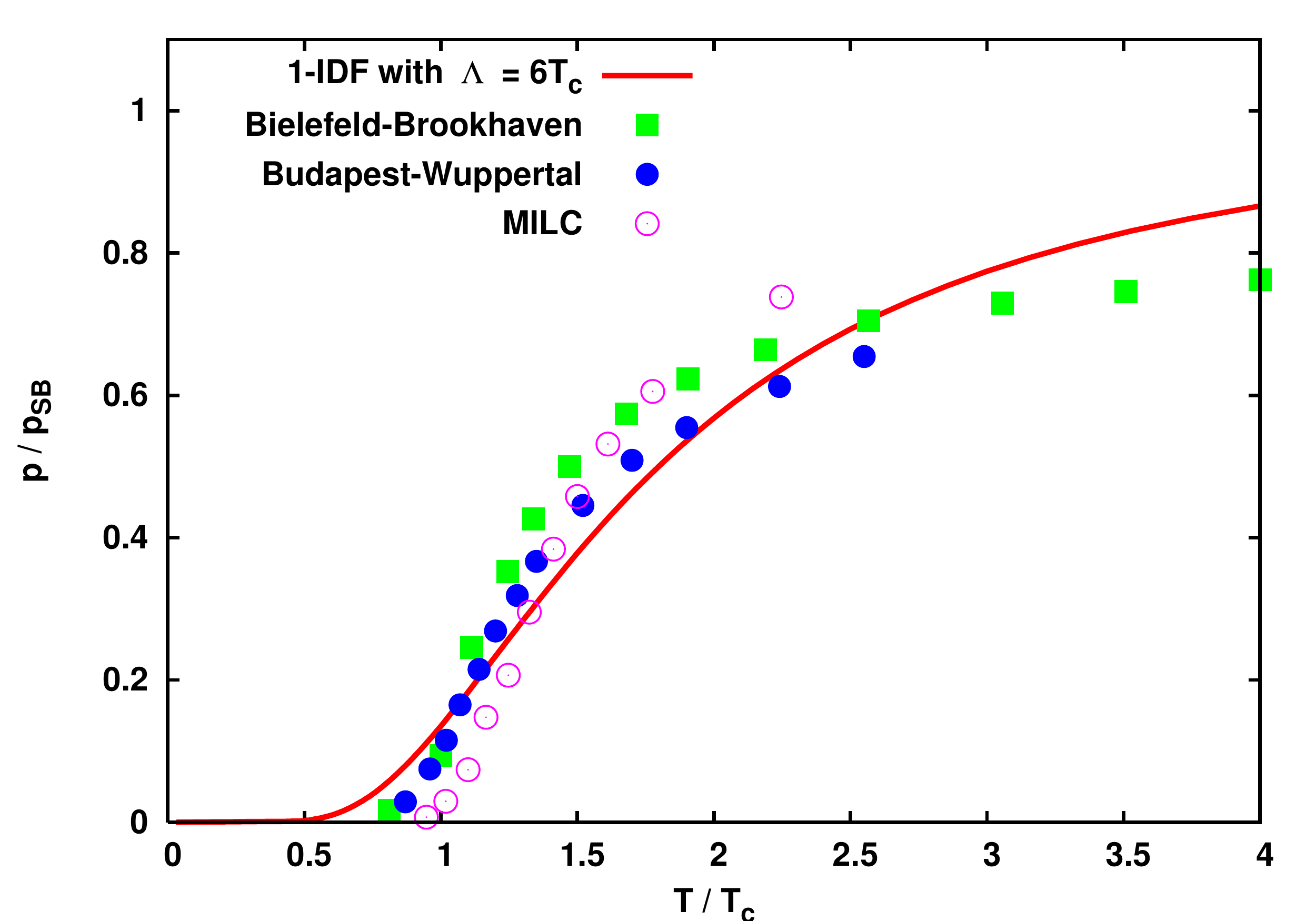}
\includegraphics[width=0.6\linewidth,angle=-90]{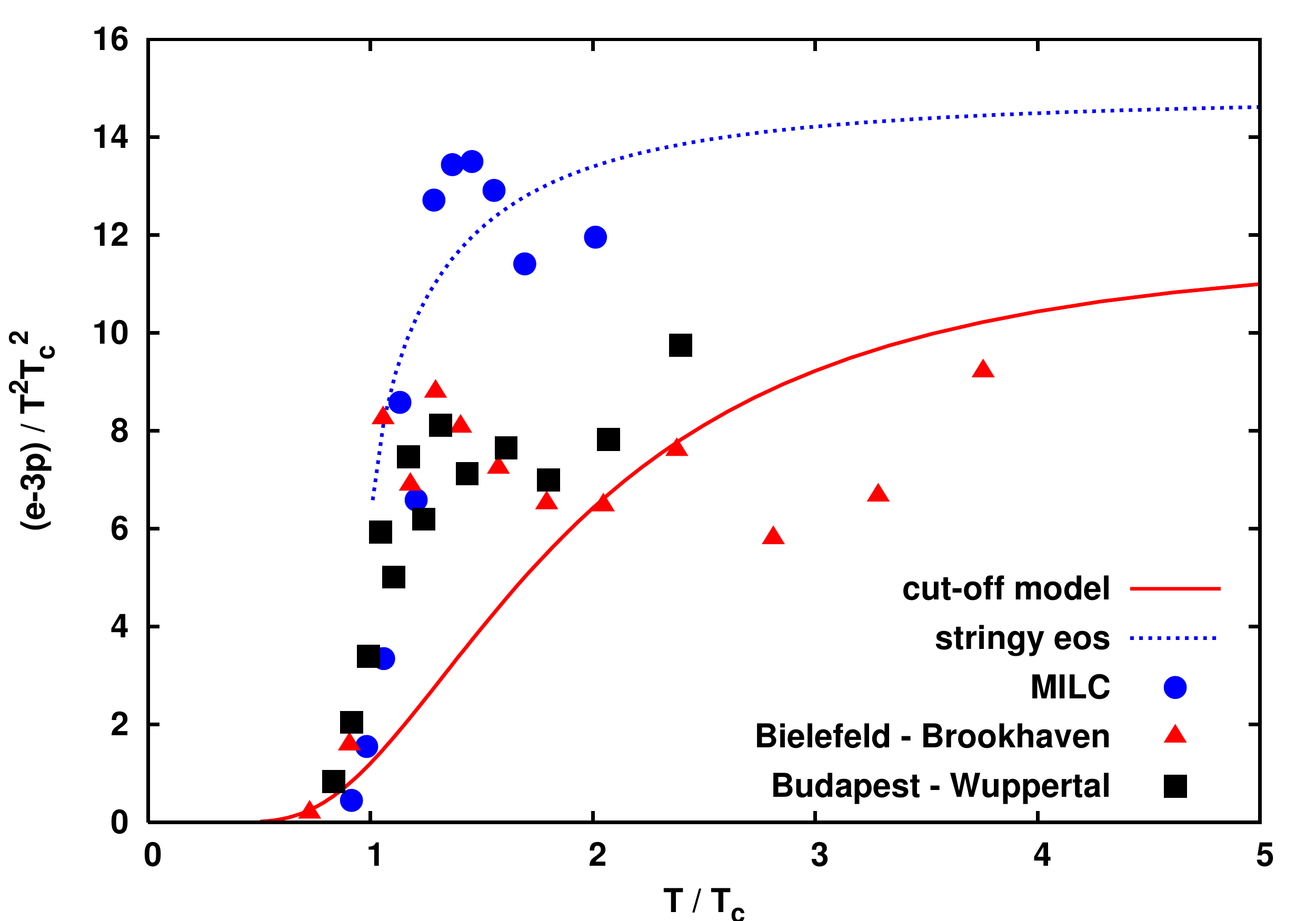}
\caption{\label{FIG3:EoSPQ2}
 Normalized pressure as the expectation value of $1-\exvs{o}$
 assuming $\Lambda=6T_c \approx 1$ GeV and lattice results (top)
 and the scaled interaction measure (cf. eq.\ref{IMPARAM}) (bottom).
 The EoS curve for stringy matter is from Ref.\cite{STRINGY3}.
}
\end{figure}

In particular for the equation of state (EoS) of high-temperature matter, among others 
for the quark-gluon plasma, NP effects are present already at this level. 
Owing to the fact that the pressure be zero in the confining phase,
one may consider that it is proportional to $1-\exvs{o}$:
\be
 p = p_{SB}\left(1-\exvs{o}\right). 
\ee{PRESS2}
Figure \ref{FIG3:EoSPQ2} plots the normalized pressure with this simple assumption
and QCD lattice equation of state data from different groups.
The cut-off parameter was taken as $\Lambda = 6T_c = 1$ GeV.
The deviation from the Stefan-Boltzmann limit is non-perturbative, showing
the subleading order at high temperature.

The interaction measure is also non-perturbative to leading order
\be
 e - 3p = 2c_{NP} \Lambda^2 T^2 + \ldots
\ee{INTMEAS}
Lattice gauge field theory calculations, especially SU(3) Yang-Mills, in fact
show a rather constant value for the parameter
\be
 \Delta = \frac{e-3p}{T^2 T_c^2},
\ee{IMPARAM}
(cf. Figure \ref{FIG3:EoSPQ2}). The perturbative QCD predicts here effects going with $T^2\alpha(T)$,
which has to be smaller than this. (The problem is that $\alpha(T) = \exvs{\alpha(Q^2)}$
also must contain NP effects, the one-loop inverse logarithm is not integrable
with the $P(Q^2)$ distribution.)


\begin{figure}
	\includegraphics[width=0.4\textwidth]{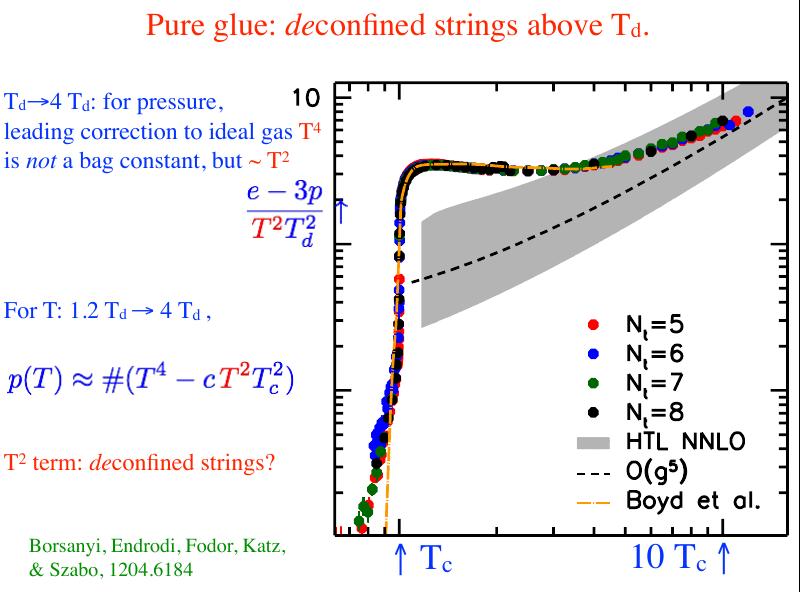}
\caption{\label{Pisarski}
	$T^2$-scaling of the interaction measure in pure glue lattice data. 
	{\em This is a slide from Rob Pisarski's lecture on July 8th, 2016, Wigner RCP Theory Seminar,
	Budapest.}
}
\end{figure}

\begin{figure}
	\includegraphics[width=0.4\textwidth]{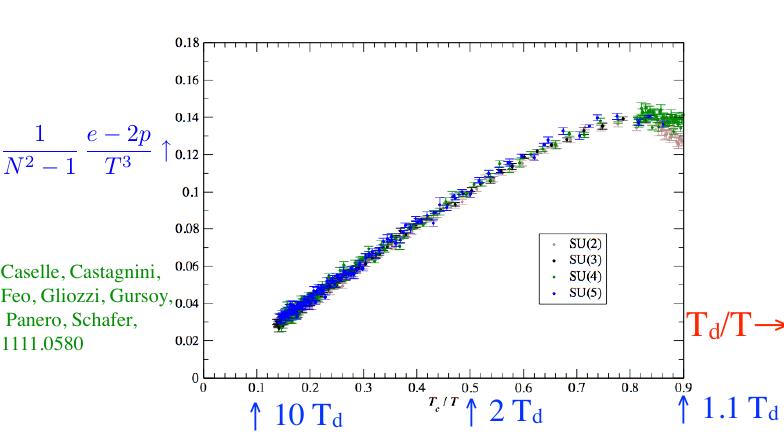}
\caption{\label{Pisarski2}
	$T^2$-scaling for $2+1$ dimensional lattice data for different SU(N) gauge groups.
	{\em This is a slide from Rob Pisarski's lecture on July 8th, 2016, Wigner RCP Theory Seminar,
	Budapest.}
}
\end{figure}


Elementary thermodynamics of ideal gases including string-like objects, gives account to this
subleading behavior. Here we briefly describe this mechanism based on the more detailed
presentation in Ref.\cite{STRINGY3}. 	
We denote the number density of colored sources by $c=\sum c_in_i$ and assume that
the temperature and number density dependent free energy density is given by
\be
 f \: = \: f_{{\rm id}}(n_i,T) \, + \, \sigma \, \exv{\ell} \,c
\ee{STRINGYf}
with $\exv{\ell}=c^{-\gamma}$. For straight color tubes with constant cross sectional
area in three dimensions one naturally assumes $\gamma=1/3$. In this way the
total free energy density contains a term depending on the total color density
as a fractional power:
\be
 f \: = \: f_{{\rm id}}(n_i,T) \, + \, \sigma \, c^{1-\gamma}.
\ee{STRINGYfc}
The chemical potentials, the pressure and energy density can be derived from this as follows
\ba
 \mu_i \: &=& \: \pd{f}{n_i} \: = \: \mu_{i,{\rm id}} \, + \, (1-\gamma) \sigma c^{-\gamma} \, c_i,
\nl
 p \: &=& \: \sum \mu_i n_i - f \: = \: p_{{\rm id}} \, - \, \gamma \sigma c^{1-\gamma},
\nl
 e \: &=& \: f - T\pd{f}{T} \: = \: e_{{\rm id}} \, + \, \sigma c^{1-\gamma}.
\ea{STRINGYeos}
Utilizing these results the interaction measure becomes
\be
 \Delta \: = \: e - 3p \: = \: e_{{\rm id}} - 3p_{{\rm id}} + (1+3\gamma) \sigma \, c^{1-\gamma}.
\ee{STRINGYim}
The contribution by the ideal gas is zero for massless objects, for each massive degree of freedom
it is proportional to $m^2$. For a QGP made of (nearly) massless quarks and gluons only the
stringy contribution remains, in which all densities are proportional to $T^3$, as $n_i=\nu_iT^3$.
In this case one obtains
\be
 \left. (e - 3p) \right|_{m=0} \: = \: 
     (1+3\gamma) \sigma \left( \sum c_i \nu_i \right)^{1-\gamma} \, T^{3(1-\gamma)}.
\ee{STRINGYmasslessIM}
For straight objects $\gamma=1/3$, and this term is proportional to $T^2$, and the same follows for the
non-ideal contribution to the total pressure, too:
\be
 p \: = \: \sigma_{SB} T^4 - \frac{1}{3} \sigma \, \left( \sum c_i \nu_i \right)^{2/3} \, T^2.
\ee{STRINGYp}
A further remarkable property of this picture is that at the edge of mechanical stability, defined by
vanishing total pressure, $p=0$, the energy density is given by
\be
 \left. e \right|_{p=0} \: = \: e_{{\rm id}} \, + \, p_{{\rm id}} / \gamma.
\ee{STRINGYe}
For massless constituents in the QGP $e_{{\rm id}} = 3 p_{{\rm id}}$, and $p_{{\rm id}} = T \sum n_i$,
and we obtain the following energy per particle
\be
 \frac{E}{N} \: = \: \frac{e}{\sum n_i} \: = \: (3+1/\gamma) T \: = \: 6T.
\ee{STRINGYEperN}
For a hadronization temperature of $167$ MeV, conjectured in earlier lattice calculations, 
this would be $E/N=1$ GeV; a value remarkably close to the result of phenomenological
fits of hadronic ideal gas mixtures (the so called ''Statistical Model'') to heavy ion experiments
at various bombarding energies\cite{STM1,STM2,STM3,STM4,STM5}. 
Since the hadronic matter has almost zero pressure compared to a QGP, the deconfinement phase
crossover transition temperature is indeed close to the mechanical instability point
defined by $p=0$.

Having a glimpse on Fig.\ref{Pisarski2} one realizes that the nonperturbative leading correction
to ideal pressure term is negative and scales with the Casimir of the charge in the $SU(N)$
gauge group, namely with $N^2-1$. Since strings pull, it is natural that they give a negative
correction to the ideal pressure. Since they are mainly made of chromoelectric flux, it is natural
that the effective string tension scales like $N^2-1$. However, for also resulting in ${\cal O}(T^2)$
corrections, in $2+1$ dimensional lattices the elementary correction per color source
must be like $f/c \sim \sigma \ln \exv{\ell}$. This hints towards a very different mechanism
for the origin of such corrections in lower dimensional Yang-Mills systems.

\vs
Summarizing this subsection, we have shown on the basis of general arguments that non-perturbative effects,
even those which cease at a sharp momentum cut-off, contribute to thermal expectation values
at arbitrary high temperatures. Based on the thermal distribution of
$Q^2$ values it was demonstrated that this contribution is of the relative order of
$\Lambda^2/T^2$ to any thermally averaged quantity.
A physical picture of such non-perturbative corrections to the ideal gas equation of state
is offered by an elementary study of the thermodynamics of straight strings with a 
naturally density-dependent average length.




\subsection{Shear viscosity bounds}
\label{sec:shear}

Accelerator experiments suggest that the matter formed in heavy ion
collisions is a very good fluid close to be a perfect one
\cite{Shuryak:2003xe,Teaney:2003kp},
which means that the characteristic dimensionless $\eta/s$ ratio
($\eta$ being the shear viscosity, $s$ the entropy density) is very
small. The quantity $\eta/s$, apart from the fact that it appears
directly in hydrodynamical formulas like the sound attenuation length
\cite{Landau}, can be interpreted as a fluidity measure of an
ultra-relativistic gas that characterizes the viscosity on its own
scale \cite{Liao:2010nv}.

The experimental tool to access this quantity is measuring the flow
anisotropy, in particular its second angular moment, $v_2$. The
desired parameters of the corresponding fluid model can be obtained by
fitting the model predictions to the experimental curves
\cite{Romatschke:2007mq,Dusling:2007gi}. The result of these studies
is that the observed $\frac\eta s \sim \frac1{4\pi}$ with a
coefficient of order one. The value $\frac1{4\pi}$ has a specific
significance, since, at is it used to say, it is the ``theoretical
lower bound''.

But, as opposed to the folklore, the status of $1/4\pi$ being a
theoretical lower bound for the $\eta/s$ ratio, is far from being
proven. We try to review in this section what are the assumptions and
approximations behind this conjecture.

The idea that $\eta/s$ can have a lower bound, was first suggested in
\cite{Danielewicz:1984ww}. The authors realized that in the kinetic
approach $\eta/s\sim E\tau$ where $E$ is the quasiparticle energy and
$\tau$ is its lifetime. For a quasiparticle $E>\Delta E$ where
$\Delta E$ is the width of the quasiparticle peak, therefore, using
uncertainty principle $\eta/s\sim E\tau > \Delta E\tau \gtrsim \hbar$,
meaning that $\eta/s$ has a lower bound. This elegant way of thought,
however, cannot be considered as a proof of the lower bound, since it
uses the kinetic, quasiparticle approach which is not really suitable
to describe the small viscosity regime. The point is that kinetic
theory estimates the shear viscosity to be $\eta\sim 1/\sigma$ where
$\sigma$ is a cross section. In weakly coupled theories
$\sigma\sim g^4$ where $g$ is the coupling constant. Since $g$ must be
small in order the kinetic, Boltzmann-equation approach be applicable,
only the large viscosity regime is accessible in this way. In this
range of applicability several studies in the literature computed the
shear viscosity using the Boltzmann-equation or quasiparticle approach
\cite{Peshier,Dobado:2009ek,Marty:2013ita,Kadam:2014cua,Bluhm:2010qf,Hidaka:2009ma}. But,
when the theory is more and more strongly coupled, higher order
processes become dominant, too \cite{Xu:2007ns,Xu:2007jv}, and the
simplest kinetic argumentation looses its validity.

One can think that perturbation theory is also can be used to
calculate the $\eta/s$ ratio. The entropy density is defined through
the thermodynamics from free energy, the shear viscosity by the Kubo
formula
\begin{equation}
  \eta =
  \lim\limits_{\omega\to0}\frac{\left\langle [T_{12}(\omega,{\bm
        k}=0),T_{12}(x=0)]\right\rangle}\omega.
\end{equation}
But with the perturbative approach there are several problems. The
first one is that in realistic applications, for example for QCD near
the critical regime of the crossover perturbation theory is not really
applicable. At somewhat higher temperatures the perturbation theory
still needs heavy machinery including resummations to reliably predict
the thermodynamical quantities like pressure or entropy density, but
after some efforts one can give a relatively good description
\cite{Andersen:2015eoa}. But the shear viscosity is a quantity that is
much harder to access. The fundamental problem is that perturbation
theory can compute corrections to a quantity calculated in the free
theory. But the shear viscosity is infinite in a free gas. Therefore
we should compute corrections to infinity which is a hard task. In a
strict diagrammatic approach one has to re-sum ladder diagrams
\cite{Jeon:1994if,Carrington:2002bv}. One can use 2PI resummation to
perform the task \cite{Aarts:2004sd}, or, concentrating only to the
most important pinch singular contributions, quantum
Boltzmann-equations
\cite{Arnold:2000dr,Arnold:2002zm,Arnold:2003zc,Jakovac:2001kj}. One
can also apply renormalization group techniques to approach the shear
viscosity \cite{Haas:2013hpa} But even after the most thorough job one
can expect a ``small'' correction to infinity that means large
numerical values: one typically gets $\eta\sim 1/(g^4\log g)$ as the
leading order estimate. For small viscosities, just like in the
kinetic approach, we would need large coupling, and so perturbation
theory is not applicable there.

An alternative approach to calculate the shear viscosity could be the
lattice Monte Carlo technique. There have been in fact attempts to
extract this information from lattice, calculating the energy-momentum
tensor correlation function \cite{Karsch:1986cq,Meyer:2007dy}. The
obtained results have been in the $\eta/s\sim 0.1-0.2$
regime. Unfortunately the measurements cannot be performed without
strong assumptions. The reason is that hydrodynamics is an effective
description of the matter only in so large timescales that is hard to
access from a Euclidean lattice. Therefore the present MC simulations
have very small sensitivity to the desired transport regime
\cite{Datta:2008dq}. We remark that in classical theories one can also
compute the shear viscosity \cite{Homor:2015qza}. Here one is not
restricted by the Euclidean formalism, but the quantum interpretation
is much more difficult.

Since small viscosity involves large couplings, therefore methods that
use the inverse coupling as expansion parameters are of great
importance. Unfortunately these dual partners are rarely
known. Therefore the conjectured AdS/CFT duality
\cite{Maldacena:1997re} has a big relevance, even though here the
weakly coupled theory is a conformal field theory, and so its
symmetries are not the same as the symmetries of QCD. Nevertheless one
can calculate the $\eta/s$ ratio in the infinitely strongly coupled
(the t'Hooft coupling $\lambda=g^2N_c$ is infinite) ${\cal N}=4$
supersymmetric Yang-Mills theory with this method \cite{Kovtun:2004de}
resulted in $\eta/s=1/{4\pi}$. The significance of this result is
raised by the fact that the infinitely coupled theory is expected to
have the smallest shear viscosity; in fact, in this model the
$1/\lambda$ corrections are all positive \cite{Myers:2008yi}. The KSS
result, together with the conjecture of the lower bound based on
kinetic approach was then advertised that ``the lower bound for the
$\eta/s$ ratio is $1/(4\pi)$''.

But, as we see, the two pivots of the argumentation are coming from
the quasiparticle and the conformal field theory limits, and so these
are not as general as it is usually thought. In fact, soon after the
announcement of the ``lower bound'' there appeared constructions that
violate, or at least challenge the $1/(4\pi)$ value. 

In the framework of non-relativistic theories one can construct such
theories \cite{Cohen:2007qr,Cherman:2007fj,Son:2007xw}; for example
theories where with the growing number of field components the shear
viscosity remains constant, but the entropy density grows with the
number of the components. It is also seems valid that when we start to
deviate from the quasiparticle approximation, for example with the
inclusion of the continuum besides the quasiparticle peak, the shear
viscosity starts to decrease
\cite{NoronhaHostler:2008ju,Jakovac:2009xn,Horvath:2015gqd}. In fact,
in very general grounds one can argue for a lower bound not for
$\eta/s$, but $\eta T^3/s^2$ 
\cite{Jakovac:2009xn,Horvath:2015gqd}.

From the Ads/CFT side, there are also doubts about the universality of
the $1/(4\pi)$ bound. Model studies of gravity models, where besides
the leading order AdS action there are corrections (higher order terms
in curvature, other fields like dilaton) lead to the conclusion that
in these models the $\eta/s$ ratio can ge below $1/(4\pi)$ 
\cite{Kats:2007mq,Buchel:2008vz,Cai:2009zv,Feng:2015oea}.
It is not clear if in a general consistent gravity model
there exists at all a lower bound.

From experimental side it seems that the $\eta/s$ ratio of the
strongly interacting plasma is ${\cal O}(1-2)/(4\pi)$
\cite{Romatschke:2007mq,Drescher:2007cd,Lacey:2013eia}. Comparing with
the $\eta/s$ values of other matters like water or even superfluid
$^4$He, the $\eta/s$ value really seems very low \cite{Lacey:2006bc},
but if we use a fluidity measure better suited for not
ultra-relativistic matter, then QCD seems to be not extraordinary
\cite{Liao:2010nv}.

So, summarizing the content of this section, although in quasiparticle
systems and some conformal theories we really expect to have a lower
bound for the shear viscosity, but in a strongly interacting matter
like QCD there is no well-established proof for that. It is also
true, that the numerical value of $1/(4\pi)$ is so small, that it is
not easy to provide such experimental setup where we could violate
this bound. But, since this bound is not a constant of nature, it can
happen that in some future collider experiments it will still be
violated.



\subsection{Rather field or rather particle?}


The particle-wave duality appears in an interesting aspect in the
heavy ion collisions. The classical picture of a particle is a
point-like object traveling on a world-line in the spacetime; if the
particle is free of interaction, the world-line is a straight line (or
geodesic line). On the other hand the free quantum particle has
infinite extension in space as a plane wave.

Interacting particles or waves penetrating and trespassing a medium get distorted.
The distortion effect depends on the nature of the interaction, its localization
and strength. Typically particle-like interactions are extremely localized,
not only in space, but also in time. The straight world-line receives kicks
once in a while. An extended medium on the other hand acts for long and makes
the particle world-line smoothly curved. The same typing for extended waves
includes changes in the dispersion relation by phase shifts in the former case and
an overall change in the latter case. Static and large media, in particular,
modify the free particle dispersion relations (propagators) by inducing a self-energy
part, which re-scales the effective mass and adds a quasi-particle width.
A continuum part in a spectral function, however, is a sign for creating and
annihilating particles during the interaction between the quantum objects and the medium.

Traditional high-temperature field theory considers the environment as given,
in most cases   keeping a sharp value of the temperature. This so called heat
bath is assumed to be god given and very few thoughts are dedicated to the problem:
where does this temperature comes from? What mechanism keeps its value so constant?
And how should we describe the QGP if none?

In principle all thermal effects are results of the same interaction.
It is therefore legitim to seek for approaches which do not assume a temperature,
but calculate it. Or at least investigate the effects due to un-sharp values of
it -- a step forward -- testing some simple distributions.
This so called superstatistical approach\cite{TSALLISorig,TSALLISBOOK,Beck1,BeckCohen,Cohen,Beck2,ABC} 
is based on a particular distribution
of $\beta$ values in the thermal weight, $\exp{(-\beta H)}$. The simplest such distribution,
having a width of $\beta$-values and allowing only non-negative ones, is an Euler--Gamma
distribution. This converts the Boltzmann--Gibbs weight in a Tsallis--Pareto form:
\be
\infi \ead{-\beta H}  \, \frac{(\beta nT)^{n-1}}{\Gamma(n+1)} \, \ead{-\beta nT} \, nT d\beta
\: = \: \left(1+\frac{H}{nT} \right)^{-n},
\ee{BG2TP}
often experienced in particle spectra measured in high-energy collisions.
Here $\exv{\beta}=1/T$ and $\Delta\beta/\exv{\beta}=1/\sqrt{n}$.
In the $n \to \infty$ limit the distribution of $\beta$ values narrows to a Dirac delta,
and the above statistical weight converts to the well-known Boltzmann factor.
Candidates for physical mechanisms therefore, which would explain the occurrence of
the temperature, $T$, in a dynamical system, should also explain whether or not the
width $\Delta\beta$ is small enough under the circumstances given.

A thermodynamical interpretation of the parameters $T$ and $n$ can be given starting
from Einstein's idea relating the statistically evenly occupied phase space volume
to the notion of entropy\cite{BiroENTROPY,BiroPHYSICA},
\be
 \Omega(E) \: = \: \ead{S(E)}.
\ee{EINSTEIN}
Picking up a subsystem with energy $H$ out of $E$ has then the probability
\be
 P(H) \: = \: \frac{\Omega(H) \, \Omega(E-H)}{\Omega(E)},
\ee{EINSTEINPROBA}
assuming no correlation other than induced by fixing the total energy to $E$.
In this microcanonical view the environmental factor,
\be
 \rho(H) \: = \: \exv{\frac{\Omega(E-H)}{\Omega(E)} },
\ee{RHOEINSTEIN}
models the statistical operator. Based on the conjectured connection to the thermodynamical
entropy one finally deals with
\be
 \rho(H) \: = \: \exv{\ead{S(E-H)-S(E)}}.
\ee{RHOEXPS}
In the expansion $H \ll E$ up to quadratic terms one obtains
\be
 \rho(H) \: = \: 1 \, - \, H \exv{S^{\prime}(E)} \, + \, \frac{1}{2}H^2 \,
 \exv{S^{\prime\prime}(E) + S^{\prime}(E)^2} \, + \, \ldots
\ee{RHOEXPANDED}
Comparing this result with eq.(\ref{BG2TP}) one interprets the parameters as
\be
 \frac{1}{T} \: = \: \exv{S^{\prime}(E)} \qquad \mathrm{and} \qquad
 T^2 \Delta \beta^2 \: = \: \frac{1}{C} + \frac{1}{n},
\ee{TSALLISINTERPRET}
with $C=dE/dT$ being the total heat capacity of the system. In the infinite reservoir (thermodynamical)
limit, $C \to \infty$, one gets back the width of the Euler--Gamma distribution.
On the other hand, for $n \to \infty$, considering a sharp $\beta$ value, one obtains the textbook
result $T^2\Delta\beta^2=1/C$ for the variance. Indeed near to the maximum the Euler--Gamma
distribution, as many other, is well approximated by a Gaussian.
However, the Gaussian assumption cannot be extended to negative $\beta$ values, therefore
the Euler--Gamma assumption is superior.

In interacting, ''real'' systems the natural dynamics itself must determine the actual
distribution of $\beta$ values. For any finite sized system, a width of this distribution
is compulsory. The task of understanding the emergence of temperature and other thermal
effects, and calculating the actual values of $T$ and $C$, and perhaps $n$ in existing
physical systems, sharpens even more in quantum (field) theory. If $\beta$, or for that
matter any function of it, is associated to an operator in the quantum description,
then its width cannot be narrowed down to zero in practice. (Only theoretically, on the cost
of having infinite width for other, non-commuting operators.)
Here the question to be answered is, how to describe a statistical operator part during
the quantum evolution, which -- under certain approximations to the physical reality --
behaves like factorizing to a Boltzmann-Gibbs, or similar, and a unitary factor.

For this purpose considering quantum states with finite width can be of help.
Contrary to point-particles (zero width in location, infinite width in momentum)
and to plane-waves (infinite width in location, zero width in momentum) more general
states, in particular coherent states look more realistic. In the rest of this section
we give a short overview of properties of coherent states and show a possible way of 
an unusual interpretation of being thermal.

\newcommand{\ket}[1]{ \left| \, {#1} \, \right> } 
\newcommand{\bra}[1]{ \left< \, {#1} \, \right| } 
\newcommand{\amp}[2]{  \left< \, {#1} \middle| {#2} \, \right> }

\newcommand{\inti}{ \int_0^{\infty}\limits\! }
\newcommand{\sumi}[1]{\sum_{{#1}=0}^{\infty}\limits\! }


\subsubsection{State labels}


We consider generalized coherent states defined by
\be
  \ket{z} = \sum_{n=0}^{\infty}\limits \sqrt{p_n(t)} \, e^{i n\Theta} \, \ket{n} 
\ee{EQ1}
with  $z = \sqrt{t} e^{i\Theta}$. 
Such constructions in quantum optics are called ''nonlinear coherent states''.
This state overlaps with the $n$-quantum state, the overlap probability being
\be
   \left|  \amp{n}{z} \right|^2 \: = \: p_n(t) \ge 0. 
\ee{EQ2}
From the normalization of the coherent state, $\ket{z}$, a normalization
of the factor $p_n(t)$ follows:
\be
 \amp{z}{z} \: = \: 
 \sum_{n,m}\limits \bra{m} \sqrt{p_mp_n} \, e^{i(n-m)\Theta}\ket{n} \: = \:
 \sum_n p_n(t) = 1.
\ee{NORM}
The expectation value of any function of the number operator,
$\hat{n}=a^{\dag}a$, is given by
\be
\bra{z} \varphi(\hat{n}) \ket{z} \: = \: \sum_n  \varphi(n) \, p_n(t).
\ee{NEXP}
This construction ensures that $p_n(t)$ is a probability distribution in ${n}$.

We also construct a complete set based on coherent states as follows:
\be
\int\!\!\frac{d^2z}{\pi}  \ket{z} \bra{z} \, = \, 
\inti\!dt \int_0^{2\pi}\limits\!\frac{d\Theta}{2\pi}  \sum_{n,m}\limits \sqrt{p_np_m}  e^{i(n-m)\Theta}
\ket{m}\bra{n}.
\ee{COMPLET1}
Here, after integrating over $\Theta$ one obtains a Kronecker $\delta_{nm}$ under the double
sum and ends up with a single sum:
\be
1 \: = \: \inti\!dt \, \sum_n p_n(t) \ket{n} \bra{n} \: = \: \sum_n \ket{n} \bra{n}.
\ee{COMPLETE}
A sufficient condition for completeness is $\inti\!dt \, p_n(t) \: = \: 1$, wishing a complete set
for all possible Fock spaces based on $\ket{n}$, it is also necessary.
This makes $p_n(t)$ to a probability distribution function of $t$ as well.
One may consider the distribution in the number of quanta $n$ as primary statistics,
while in the coherent state parameter $t=|z|^2$ as superstatistics.


\subsubsection{Operator eigenstate}

While the most known coherent states show a Poisson statistics in $n$,
and are eigenstates of the annihilation operator, $a$, the generalized
versions are eigenstates to a more complex operator.
To construct this operator is related to the problem of regularizing
the phase operator in quantum optics \cite{PHASEOP1,PHASEOP2,PHASEOP3,PHASEOP4}.	

We request that $\ket{z}$, defined as in eq.(\ref{EQ1}), is an eigenstate 
with eigenvalue $z$ to the following operator:
\be
F \ket{z} \: = \: a g(\hat{n}) \ket{z} \: = \: z \, \ket{z}.
\ee{OPER}
Here $a$ is an annihilating ($a^{\dag}$ is a creating) operator,
and $\hat{n} = a^{\dag}a$ is the number operator. $g(\hat{n})$ is a yet
unspecified function of the number operator.

The action of this operator $F$ on the general coherent state causes
\be
 F \ket{z} \: = \:
 \sum_{n=1}^{\infty}\limits g(n) \sqrt{np_n} \, e^{in\Theta} \ket{n-1}, 
\ee{F_ACTS}
that can be re-indexed to
\be
 F \ket{z} \: = \:
 \sum_{n=0}^{\infty}\limits g(n+1) \sqrt{(n+1)p_{n+1}} \, e^{i(n+1)\Theta} \ket{n}.
\ee{F_ACTS_SHIFTED}
One can derive a recursion law by comparing this result with
\be
 z \, \ket{z} \: = \: \sqrt{t} \, e^{i\Theta} \sum_{n=0}^{\infty}\limits
 \sqrt{p_n} \, e^{in\Theta} \ket{n}.
\ee{Z_ACTS}
In conclusion the following relation has to be satisfied:
\be
p_n(t) \: = \: \frac{t}{n g(n)^2} \, p_{n-1}(t).
\ee{RECUR}
This specifies to a known quantum number distribution, $p_n(t)$, the necessary
function $g(n)$, or serves as a recursion rule for a given $g(n)$, fixing the
operator $F$, to obtain the distribution $p_n(t)$.
The recursion is solved by
\be
p_n(t) \: = \: p_0(t) \, \frac{t^n}{n!} \: {\prod_{j=1}^{n}\limits g(j)^{-2}}.
\ee{PN_SOLV}
Finally $p_0(t)$ can be obtained from the normalization condition.
At the end of this reconstruction also the completeness constraint, 
$\inti\!dt \, p_n(t) \: = \: 1$,  has to be checked.


\subsubsection{Glauber and Negative Binomial states}
The most known, traditional coherent state is defined by $\bm{g(n)=1}$.
This results in a {{ Poisson}} distribution  in $n$ 
and in an {{ Euler-Gamma}} distribution in $t$:
\be
 p_n(t) \: = \: \frac{t^n}{n!} \, e^{-t}.
\ee{POISSON}
In this case $\ket{z}$ is an eigenstate to the $F=a$ annihilation operator.

The negative binomial coherent state is based on the
negative binomial distribution (NBD),
\be
p_n(t) \: = \: \binom{n+k}{n} \, (t/k)^n \, (1+t/k)^{-n-k-1}.
\ee{NBD}
It is a normalized NBD in $n$, and at the same time is an Euler-Beta distribution in  $t$.
From the recursion eq.(\ref{RECUR}) one obtains the necessary $g(n)$ function for modifying
the operator $a$ to $F=ag(\hat{n})$:
\be
g(n)^2 \: = \: \frac{t}{n} \, \frac{p_{n-1}}{p_n}
  \: = \: \frac{k+t}{k+n},
\ee{FN2}
so this NB coherent state satisfies
\be
a \sqrt{\frac{k+|z|^2}{k+a^{\dag}a}} \, \ket{z} \: = \: z \, \ket{z}.
\ee{NBD_CS}

One realizes a one-dimensional boost property beyond such NB states when introducing
the rapidity-like notation: $t/k = \sinh^2\zeta$. Using this notation the distribution
and the Fock-representation of such a state are rewritten as
\ba
&&p_n(t) = \binom{k+n}{n} \, \sinh^{2n}\zeta \, \cosh^{-2n-2k-2}\zeta,
\nl 
&&\ket{z} =  
\sumi{n} \sqrt{\binom{k+n}{n}} \, \frac{\left(e^{i\Theta} \, \tanh\zeta\right)^n}{\cosh^{k+1}\zeta}\ket{n}.
\ea{NBD_ZETA}
Using the velocity variable $v=\tanh\zeta$, the corresponding Lorentz factor 
is given by $\gamma=\cosh\zeta$ and the overlap probability between two 
NB coherent states becomes
\be
\left| \amp{z_1}{z_2} \right|^2 \: = \:
\left[1 +\gamma_1^2\gamma_2^2 \left| v_1 e^{i\Theta_1} - v_2 e^{i\Theta_2} \right|^2 \right]^{-k-1}.
\ee{OVERLAP}
This result reminds us to a Tsallis--Pareto distribution
with the energy variable replaced by a relative velocity squared
in a 2+1 dimensional vector notation. The possibility of using such a notation is
related to the $su(1,1)$ algebra structure of operators forming an NB coherent state,
an interesting connection which shall be discussed below.

For this purpose it is enlightening to express the overlap between NB coherent states
in terms of the complex numbers $z_1$ and $z_2$.
We introduce the following vector,
\be
\vec{K} = \gamma_1\gamma_2 \left( \vec{v}_1 - \vec{v}_2 \right) 
\: = \: \frac{1}{\sqrt{k}} \, \left( \gamma_2 z_1 - \gamma_1 z_2 \right).
\ee{KDIFF}
The NB state overlap written this way converges to the known overlap
between Glauber coherent states for large $k \to \infty$:
\be
\left| \amp{z_1}{z_2} \right|^2 \: = \:
\left[ 1 + \frac{1}{k} \left|\gamma_2 z_1 - \gamma_1 z_2 \right|^2 \right]^{-k-1}
\to e^{-|z_1-z_2|^2}.
\ee{Z_OVER}
with $\gamma_i = \sqrt{1+|z_i|^2/k} \to 1$.
It is an interesting question whether we can connect some physical property of
particle-like objects to this quantity. Interpreting the $v_i$-s as magnitudes
of velocities and obtaining the corresponding Lorentz factors from the energy
to mass ratio of pointlike massive objects one would consider
\be
\vec{K} = \frac{1}{m_1m_2} \left( E_2 \vec{P}_1 - E_1 \vec{P_2}  \right).
\ee{K_VEC_PART}
In this interpretation the overlap between two NB states decays asymptotically as a power-law
of the relative velocity of relativistic massive particles moving on a plane.


It is useful to extend the notation of NB states with the index $k$, referring to
the $k$ parameter in the underlying NBD distribution.
From now on we use the following notation
\be
\ket{ z_k, k } \: = \: \sum_{n=0}^{\infty}\limits \sqrt{p_n(k)} e^{i n\Theta} \ket{n} 
\ee{OWN_NBS_DEF}
with $z_k=\sqrt{kf}\, e^{i \Theta}$ for the NB states and 
\be
 p_n(k) \: = \: \binom{k+n}{n} \, f^{n} \, (1+f)^{-n-k-1}
\ee{OWN_NBS_COEFF}
for the NBD distribution. This distribution
provides an average number of $\exv{n} = (k+1)f$.

We are primarily interested in the effect of annihilating a quantum from an NB state.
The annihilating ladder operator acts as
\be
a \ket{z_k,k} \: = \: 
 \sum_{n=1}^{\infty}\limits \sqrt{p_n} e^{in\Theta} \, \sqrt{n} \ket{n-1}
\ee{OWN_A_NB1}
what can be re-indexed into
\be
a \ket{z_k,k} \: = \: 
 \sum_{n=0}^{\infty}\limits \sqrt{(n+1)p_{n+1}} \, e^{i\Theta} \, e^{in\Theta} \ket{n}.
\ee{OWN_A_NB0}
Consider now that
\be
(n+1) p_{n+1}(k) \: = \: 
(n+1) \binom{k+n+1}{n+1} f^{n+1} (1+f)^{-(n+1)-k-1} 
\ee{OWN_NpN}
can be expressed as
\be
(n+1) p_{n+1}(k) \: = \: 
f({k+1})  \, \binom{k+1+n}{k+1} f^{n} (1+f)^{-n-(k+1)-1}.
\ee{OWN_Npk}
Here we recognize $z_{k+1}=\sqrt{f(k+1)} e^{i\Theta}$ as a factor and arrive at
the elegant result
\be
a \, \ket{z_k, k} \: = \: z_{k+1} \ket{z_{k+1}, k+1}.
\ee{OWN_NBS_STEP2}
In the $k\to\infty$ limit again the familiar Glauber coherent state emerges, as an eigenstate
of the annihilation operator, $a$. On the other hand for an NB state, the annihilation
of a primary quantum can be represented by an upgrade of the parameter $k$ by one, times the
corresponding complex eigenvalue, $z_{k+1}$.
This helps us to answer the question of to what operator is an NB state an eigenstate.
We compare the above result with the action of another operator 
\be
\sqrt{\hat{n}+k+1} \, \ket{z_k, k} \: = \: \sqrt{(k+1)(1+f)} \, \ket{z_{k+1},k+1},
\ee{NBS_SQRT_ACTION}
based on the relation
\be
(k+1+n) \, \binom{k+n}{k} \: = \: (k+1) \,\binom{k+1+n}{k+1}.
\ee{OWN_BINOM_STEP}
This helps to recognize the following NB eigenvalue equation:
\be
\left(\sqrt{f} \, (\hat{n}+k+1) -\sqrt{1+f} \, e^{-i\Theta} \, \sqrt{\hat{n}+k+1} \: a \right) \, \ket{z_k, k} \: = \: 0.
\ee{OWN_EIGENV}
Based on this it is now easy to work out the corresponding NB algebra.
In the previous eigenvalue equation (\ref{OWN_EIGENV}) the following operator occurs:
\be
 K_- \: = \: \sqrt{\hat{n}+k+1} \, a, \qquad  
 K_+ \: = \: K_-^{\dag} \: = \: a^{\dag} \, \sqrt{\hat{n}+k+1}. 
\ee{KALGEBRA}
Their commutator,
\be
\left[ K_-, K_+ \right] \: = \: (\hat{n}+1)(\hat{n}+k+1) - \hat{n}(\hat{n}+k)
\: = \: 2\hat{n}+k+1 \: = \: 2K_0
\ee{OWN_COMMUT}
defines $ K_0 \: = \: \hat{n} + (k+1)/2$ as a linear expression using $\hat{n}$. 
With $\alpha = \sqrt{(1+f)/f} \, e^{-i\Theta}$ we arrive at
\be
\left(\alpha \, K_- -  \, K_0 \right) \, \ket{z, k} \: = \: \frac{k+1}{2} \cdot \ket{z,k}.
\ee{OWN_NBS_EIGENV2}
The commutators among the $K$-operators form an SU(1,1) algebra:
\ba
 \left[ K_0, K_+ \right] \: &=& \: K_+
\nl
 \left[ K_0, K_- \right] \: &=& \: -K_-
\nl
 \left[ K_-, K_+ \right] \: &=& \: 2K_0
\ea{COMS}
The Casimir operator is given as: $Q = K_0^2 - K_0 - K_+K_-.$

\subsubsection{Creation of an NB state from Fock vacuum}


Another important question is how to create an NB state from the vacuum.
For the ordinary coherent state a unitary operator, also called Weyl operator,
does the job. For a more general analysis we consider again general functions
of the number operator, $\hat{n}$, mixed with annihilation and creation operators.
This describes a coupling to a field with field-energy dependent coefficients.

The known operator identity,
\be
e^{A+B} \: = \: e^{-\lambda/2} \, e^{A} \, e^{B} \qquad {\mathrm{with}} \qquad
[ A, B ] \: = \: \lambda,
\ee{OPIDENT}
if $[A,\lambda]=0$ and $[B,\lambda]=0$,
helps us to obtain the necessary form for $\ket{z_k,k}=U\ket{0}$.
We choose \quad $A = \alpha \, z f(\hat{n}) \, a^{\dag}$  \quad and 
\quad $B = - \beta \, z^* a  \, / f(\hat{n})$. 
Please note that $ B \ne - A^{\dag}$. 
The commutator,
\be
[A,B] = |z|^2 \alpha \beta \, 
\left( a \frac{1}{f(\hat{n})} f(\hat{n}) a^{\dag} - f(\hat{n}) a^{\dag}a \frac{1}{f(\hat{n})} \right)
\ee{COMMUTATOR}
leads to $\lambda = |z|^2 \alpha \beta$.
We seek our evolution operator from the Fock vacuum to an NB state in the form
\be
U \: = \: e^{\Phi/2+A+B} \: = \: e^{(\Phi-\lambda)/2} \, e^A \, e^B.
\ee{EVOLU}
Here $e^B\ket{0}=\ket{0}$ due to $B\ket{0}=0$, since $B$ in the exponent
is proportional to the annihilation operator.  Expanding the exponential one obtains
\be
U\ket{0} \: = \: e^{(\Phi-\alpha\beta \, |z|^2)/2} \, \sumi{n} 
\frac{\alpha^n z^n}{n!} \left(f(\hat{n}) a^{\dag} \right)^n \, \ket{0}.
\ee{UONVAC}
Since we have
\be
 \left(f(\hat{n}) a^{\dag} \right)^n \, \ket{0} \: = \:
 f(n)\cdot \ldots \cdot f(1) \, \sqrt{n!} \, \ket{n},
\ee{NPHOTONSTATE}
the form
\be
 U \ket{0} \: = \: \sumi{n} \sqrt{u_n} \, e^{in\Theta} \ket{n},
\ee{UFORM}
with $z=\sqrt{t}\, e^{i\Theta}$, is achieved when 
using $f(\hat{n}) = \xi \, \sqrt{\hat{n}+k}$:
\be
u_n = e^{\Phi-\alpha\beta\, t} \, \left(\alpha^2 \xi^2 t \right)^n \, \binom{n+k}{n}.
\ee{UCOEFF}
In order to satisfy normalization of the resulting NB state, $||U\ket{0}||^2=1$, one needs
\be
 \sumi{n} u_n \: = \: e^{\Phi-\alpha\beta t} \, \left(1-\alpha^2\xi^2t \right)^{-(k+1)} \: = \: 1.
\ee{USUM1}
From this we express
\be
\alpha^2\xi^2t = 1 - e^{\frac{1}{k+1}(\Phi-\alpha\beta t)} \: = \: 1-w,
\ee{AXI2}
and gain the following negative binomial distribution:
\be
  u_n \: = \: \binom{n+k}{n} \, w^{k+1} \, (1-w)^n.
\ee{UNBD}
For the superstatistics normalization, we utilize the Euler Beta integral
\be
 \inti dt \, u_n(t) \: = \: 
 \binom{k+n}{n} \, \inti dt \, w^{k+1} (1-w)^n, 
\ee{SUPERUT}
which after changing the integration variable from $t$ to $w$ should become unity:
\be
 \inti dt \, u_n(t) \: = \: 
 \binom{k+n}{n} \, k \, \int_0^1\limits dw \, w^{k-1} (1-w)^n \: = \: 1.
\ee{SUPERUT2}
This condition is satisfied if
\be
 dt = -k\frac{dw}{w^2}, \qquad \longrightarrow \qquad
 t \: = \: k \, \left(\frac{1}{w} - 1 \right),
\ee{TASOFW}
or -- expressing $w(t)$ -- the superstatistics is normalized if
\be
w \: = \: \frac{1}{1+t/k} \: = \: \frac{k}{t+k}.
\ee{WASOFT}
This requirement identifies $\Phi$ as being
\be
\Phi \: = \: \alpha \beta \, t - (k+1) \ln \left( 1 + t/k \right)
\ee{PHIEXPRESSED}
After these manipulations only $\alpha, \beta$ and $\xi$ remain undetermined.
A purposeful choice is $\alpha = \beta = 1/\sqrt{t} = 1/|z|$ with
$\xi=1/\sqrt{t+k}$.
The natural logarithm of the evolution operator becomes in this case
\ba
	\ln U \: &=& \: - \frac{k+1}{2} \ln \left(1 + t/k \right) + \frac{1}{2}
\nl
& &	- \, e^{-i\Theta} \, a \, \sqrt{\frac{t+k}{\hat{n}+k}} 
	+ \, e^{i\Theta} \sqrt{\frac{\hat{n}+k}{t+k}} \, a^{\dag}.
\ea{LOGUNBD}
This is {{\em neither Hermitean nor anti-Hermitean}}, therefore can only be interpreted
as 
\be
 \ln U \: = \: - \frac{1}{2} \, \beta H_0  +  \frac{i}{\hbar}\tau H_1.
\ee{LOGUASSUM}
With this definition one has a statistical operator of
\be
 \rho \: = \: U^{\dag}U \: = \: e^{-\beta H_0}.
\ee{RHOBYLOGAU}
This underlines the necessity of thinking in complex time paths or equivalently
assuming an environmental factor, which is not unitary.
The anti-Hermitean part of $\ln U$ gives a guess for the evolution Hamiltonian
\be
 \frac{i}{\hbar} \tau H_1 \: = \: \ead{i\Theta} f_+(\hat{n})a^{\dag} - \ead{-i\Theta} a f_+(\hat{n}),
\ee{ANTIHERMPART}
while the Hermitean part for the statistical factor
\be
 -\frac{1}{2} \beta H_0 \: = \:  R_k \, + \, \ead{i\Theta} f_-(\hat{n}) a^{\dag} + \ead{-i\Theta} a f_-(\hat{n}).
\ee{HERMPART}
Here
\ba
 f_{\pm}  \: &=& \: \frac{1}{2} \left(\sqrt{\frac{\hat{n}+k}{t+k}} \pm \sqrt{\frac{t+k}{\hat{n}+k}} \right),
\nl
 R_k \: &=& \: \frac{1}{2} - \frac{k+1}{2} \ln (1+t/k).
\ea{RDEFFDEF}
Please note that in this case the statistical environmental operator contains a factor in the
Tsallis--Pareto form:
\be
 \rho \: \propto \: \left( 1 + \frac{t}{k} \right)^{-k-1},
\ee{RHOTSALLIS}
with $t=|z|^2$ interpretable as the average number of quanta in an NB state.

\subsection{Physical sources of NB states}

Various physical mechanisms may lead to an NBD in particle numbers produced in high energy
collisions. The NB coherent state, reviewed above, is just one of them; more closely we did
not specify the Hamiltonian which produces such a state. Typically, as described in
eqs.(\ref{ANTIHERMPART}) and (\ref{HERMPART}), a linear field coupling of the particle modes,
as described by the phase factors $\ead{\pm i \Theta}$ and the annihilation and creation
operator, are necessary for the desired result. However, other than for the classical Glauber
coherent state, the coupling (or equivalently the amplitude of the external field
creating the particles) does depend on the number of particles to be produced.
In particular for the NB coherent state this dependence is algebraic, involving the
square root function -- as specified in eq.(\ref{RDEFFDEF}). This effect is typically
a multiparticle effect, the mathematical expression involves infinitely many quanta
for the same mode.

There are further hints and speculations about the origin of negative binomial particle
distributions. A simple phase space cell statistics, ''throwing'' stones of quanta into
phase space cells repeatedly in unrestricted number -- as bosons behave -- predicts
a combinatoric factor in the probability to find exactly $n$ particles in $k$ cells.
The P\'olya distribution extends this picture to the same probability, when the investigated
system of $n$ and $k$ is a part of a bigger system consisting of $N$ particles altogether
in $K$ total number of phase space cells:
\be
	P_n \: = \: \frac{\binom{k+n}{n} \, \binom{K-k+N-n}{N-n} }{ \binom{K+N+1}{N} }.
\ee{POLYA}
In the large environment limit, $K\to\infty, N\to\infty$  while  $f=N/K$ is kept finite,
one arrives at the NBD distribution (\ref{OWN_NBS_COEFF}).
This approach predicts that $k+1 = \exv{n}/ \, f \propto N_{{\rm part}}$, 
so high-multiplicity events are closer to the Poissonian, if $f$ is universal.

In the derivation of an NB state in quantum optics, the $f$ parameter occurs
as a squeeze parameter \cite{PHASEOP4,Jackiw}.  
Analyzing bosonic wave packet statistics,
onefold filled bosonic states give a correlation factor
of $2$ at zero relative momentum. With $M$-fold occupation of the same
state it reduces to
\be
 C_2(0) \: = \: 1 + \frac{1}{M} \: = \: 1 + \frac{1}{k+1}.
\ee{HBT_ZERO}
The logarithmic cumulants for an NB state, defined by $G(z)=\sum p_n z^n $ and
$\ln G(z) = \sum_{n=1}^{\infty}\limits C_n(z^n-1)$, are
\be
C_n = \frac{k+1}{n} \left( \frac{f}{1+f} \right)^n. 
\ee{NBS_LOGCUM}
This looks like a $(k+1)$-fold overload of the simple Bose case, given if $k=0$.
\cite{KauffmannGyulassy,CsorgoPrattZimanyi}  

Regarding the superstatistics view \cite{Beck1,BeckCohen,Cohen}, 
thermodynamical $\beta$-fluctuations and $n$-fluctuations are related by
the Poisson transform. From
\be
\inti \gamma(\beta) e^{-\beta\omega} d\beta \: = \: 
\sumi{n}  \left(1-\frac{\omega}{E} \right)^n \, P_n(E).
\ee{POITRF}
it follows
\be
 P_n(E) \: = \: \inti \frac{(\beta E)^n}{n!} \, \ead{-\beta E} \, \gamma(\beta) \, d\beta,
\ee{POITRFEXPANDED}
after Taylor-expanding $\ead{\beta E(1-\omega/E)}$ 
in the factorized expression $\ead{-\beta\omega}=\ead{-\beta E}\ead{\beta E (1-\omega/E}$.

\noindent
In this way $\Delta\beta^2/\exv{\beta}^2 = 1/(k+1)$ cf. eq.(\ref{TSALLISINTERPRET}). 

In perturbative QCD calculations \cite{Dokshitzer,Dremin,DD2} 
KNO scaling + DGLAP give nearly NBD with constant $k$, related to
$\Lambda_{QCD}$ and expressed by the $n$-variance.
In experimental findings NBD is in fact slightly violated, so more sophisticated
distributions are also discussed.

In a minimal information statistical approach, the Tsal\-lis--Pareto distribution
in stead of the Boltzmann--Gibbs is viewed as a maximal entropy state.
Utilizing the Tsallis' or R\'enyi entropy formula \cite{TSALLISBOOK,Renyi} 
the usual canonical constraint on the average energy indeed leads to 
\be
w(\omega) = \left(1 + (q-1) \frac{\omega}{T} \right)^{-\frac{1}{q-1}}.
\ee{TSALLISPARETO}
Using further assumptions about reservoir fluctuations, further entropy formulas can be
constructed, as expectation values of formal logarithms, behaving additively
\cite{Biro:Entropy2016}.

Finally we have to mention the color glass condensate picture  \cite{CGCA,CGCB,CGCC,CGCD}
where particles are produced directly from the decay of a nearly classical non-Abelian field.
A glittering glasma with $k$-fold ropes delivers also NBD in the final state.
In this model the main NBD parameter, $k$ is calculated as
$k=\kappa (N_c^2-1) Q_s^2 R^2 / 2$, with $\kappa$ string constant,
$N_c$ number of colors, $Q_s$ saturation energy scale and
$R$ radius in transverse plane. Its value is about the number of tubes, produces NBD
with this parameter. The actual estimates in Ref.\cite{GelisLappi} go
to a few hundreds for heavy ion collisions at RHIC.



\renewcommand{\d}{\partial}
\renewcommand{\c}[1]{{\cal{#1}}}

\renewcommand{\Im}{\, \mathfrak{Im}\, }
\renewcommand{\Re}{\, \mathfrak{Re}\, }

\subsection{Lattice field theory with canonical Tsallis distribution}

Lattice field theory is still the only systematic nonperturbative 
computational tool to solve physical problems related to the 
strong interaction. While at its dawn it produced only
a qualitative insight into the characteristic features of the strongly
interacting matter, in the last decades it has accomplished a lot and
produced reliable quantitative predictions on equilibrium quantities
such as the transition temperature or the equation of state.
Among them, for example,  a precise study of the transition 
temperature has been done with staggered fermions and results from
different groups tend to agree quite remarkably \cite{BW10,HQCD11}.
It has been shown that the deconfining transition is a crossover
for physical quark masses \cite{Aoki06} and the deconfining transition 
and the chiral transition essentially coincide.

Despite the remarkable success of lattice calculations regarding 
equilibrium quantities, a direct investigation of out-of-equilibrium
properties is not possible using conventional lattice gauge theory
methods. There are attempts to describe near-equilibrium 
properties, especially via spectral functions. For a more detailed overview 
we direct the reader to a recent review \cite{Meyer15} on this topic.

Here we present a completely different approach within the lattice 
gauge theory framework, which is based on the generalized, non-extensive
thermodynamics. Non-extensive thermodynamics is regarded as an 
effective theory for non-equilibrium effects and long-range correlations
\cite{Tsallis09}. One possible manifestation of it in the language of
probability distributions is the Tsallis distribution, 
\begin{equation}
 w_i(E) = \frac{1}{Z_{TS}} \: \left( 1 + \frac{\beta E_i}{c} \right)^{-c}
\label{TS-DIST}
\end{equation}
where $E$ denotes the energy of a state and 
$\beta$ is the inverse temperature. This power-like function 
restores the Gibbs factor in the $c \rightarrow\infty$ limit:
\begin{equation}
 \lim_{c\rightarrow\infty} w(E) = \frac{1}{Z_G} \exp(-\beta E)
\label{GIBBS-DIST}
\end{equation}
Observed particle spectra in high energy $pp$ and heavy-ion reactions can be well described
with the above distribution \cite{Urmossy,Urmossy2}.
Via the Gamma-distribution
\begin{equation}
 w_c(\theta) = \frac{c^c}{\Gamma(c)} \, \theta^{c-1} \, e^{-c\theta}
\label{GAMMA-DIST}
\end{equation}
the Tsallis weight can be rewritten as
\begin{equation}
 w_i = \frac{1}{Z_{TS}} \, \int_0^{\infty}\!d\theta \, 
 w_c(\theta) \exp(- \theta \beta E_i), 
\label{SUPERSTAT}
\end{equation} 
which is a specific case of the superstatistical approach
\cite{SUP1,SUP2}. That means that averages with the Tsallis distribution
can be calculated from the Gibbs expectation values at different $\beta$'s:
one simply has to average them over the inverse temperature, $\beta$, 
which obeys a Gamma-distribution. The corresponding partition functions
are connected as
\begin{equation}
\begin{split}
 Z_{TS}(\beta) &= \sum_i \int_0^{\infty}\!d\theta \, w_c(\theta) 
 \exp(-\theta \beta E_i) \\
  &=  \int_0^{\infty}\!d\theta \, w_c(\theta)  Z_G(\theta \beta).
\end{split}
\label{ZTS-ZG}
\end{equation}

As a consequence, so as to take into account non-equilibrium effects,
it is possible to apply the Tsallis distribution in lattice gauge theory 
simulations instead of the Gibbs one. The calculation of expectation values
can be realized using the superstatistical method.  
For an exploratory study, the simplest, pure SU(2) theory has been chosen
and investigated so far \cite{BiroSchram1,BiroSchram2,BiroSchram3}.  

In a lattice simulation, the physical temperature is determined 
by the period length in the Euclidean time direction: $\beta = N_ta_t$.
It is clear that one cannot use the re-sampling technique of the 
traditional, Gibbs distributed configurations in order to evaluate 
the Tsallis averages. Those configurations are available for a few integer
$N_t$-s only, therefore the necessary coverage of the Gamma distribution
is not possible. However, if the timelike lattice spacing, $a_t$ follows a 
Gamma distribution, so do the physical $\beta = N_ta_t=1/T$ inverse temperature
values at a given, fixed $N_t$. We assume that the mean value $< a_t >$
is equal to the spacelike lattice spacing, $a_s$. 
Then the ratio $\theta=a_t/a_s$ follows
a normalized Gamma distribution with the mean value $1$ and a width of
$1/\sqrt{c}$. Inspecting ZEUS $e^+e^-$ data we obtain $c \approx 5.8 \pm 0.5$. 
In the numerical calculations outlined next the value $c=5.5$ has been used. 

The form of equation (\ref{ZTS-ZG}) for gauge fields is given as
\begin{equation}
  Z_{TS}\left[\beta\right] \, = \,
  \int_0^{\infty}\!d\theta \, w_c(\theta)
  \int {\cal D}U \, e^{-S\left[U,\theta \right]}  
 \label{PATHINT}
\end{equation}
In general, one needs to calculate  
the Tsallis expectation value of an observable $\hat{A}[U]$ over lattice field
configurations $U$. If $\hat{A}$ has a form $\hat{A}=\theta^{\:v}A$,
one obtains
\begin{equation}
 \langle A \rangle_{TS} \, =  \, \frac{1}{Z_{TS}} \frac{c^c}{\Gamma(c)}
 \int\!d\theta\: \theta^{\: c-1} e^{-c\:\theta} \int {\cal D}U A\left[U\right]
 \theta^{\:v} e^{-S\left[\theta,U\right]}
\label{TS-EXP}
\end{equation}
Here the lattice action generally can be written as
\begin{equation}
 S\left[\theta,U\right] = a \: \theta + b / \theta,
 \label{SLAT}
\end{equation}
where 
$a=S_{ss}[U]$ contains space-space oriented plaquettes and
$b=S_{ts}[U]$ contains time-space oriented plaquettes.
In the 
$c \rightarrow \infty$ limit the Gamma distribution approximates
$\delta(\theta-1)$, and one gets back the traditional lattice
action $S=a+b$, and the corresponding averages. For a given
finite $c$, one can exchange the $\theta$ integration and the 
path integral and obtains exactly the power-law-weighted expression.

One can indeed explicitly perform the $\theta$ integration and 
derive an effective action, which turns out to be a logarithm 
of a Bessel function. As it is not particularly easy to carry on
simulations with that effective action, we consider an other 
solution here.

According to equation (\ref{TS-EXP}), it is possible to modify the direct 
numerical update algorithms used in the usual canonical Gibbs simulations 
for our superstatistical ensemble. We use the Metropolis algorithm
here, but the heat-bath should also work for pure gauge fields. 
Now there is an additional variable, the anisotropy,
which has to be generated according to a Gamma distribution with the 
Tsallis parameter $c$. 
Simulations for different $c$ values has been performed in the range
of 5.5 -- 1024.0, the lower value being relevant  
in high-energy collision experiments, while
$c = 1024$ is meant to approximate the  $c \rightarrow \infty$ (Gibbs) 
limit.
At a given $c$ the numerically generated random $\theta$ anisotropy values
show, that for a really smooth reconstruction of the
Euler-Gamma distribution one needs random values in the order 
of $10^4 \sim 10^5$. Of course, this criterion considerably elongates
the simulations and makes them computationally more expensive.

Once an anisotropy parameter is given, the standard Metropolis 
sweep is applied to update the gauge fields. After a sweep through the whole 
lattice a new $\theta$ is thrown and so on.
The ensemble averages are calculated with averaging the gauge configurations 
over the generated $\theta$ values in the standard way.
The numerical calculations show that the - here outlined - generalized Monte 
Carlo method is stable and functioning. Simulations have been performed
on various lattices up to $10^3 \times 2$ and $10^4$ sizes and different
thermodynamical quantities, including the equation of state, have been
estimated \cite{BiroSchram1,BiroSchram2} via the calculation used
in the pioneering work of \cite{Engels81}. The $c=1024$ results can 
also be compared directly to those obtained in \cite{Engels81}.

\begin{figure}
\includegraphics[width=0.9\linewidth]{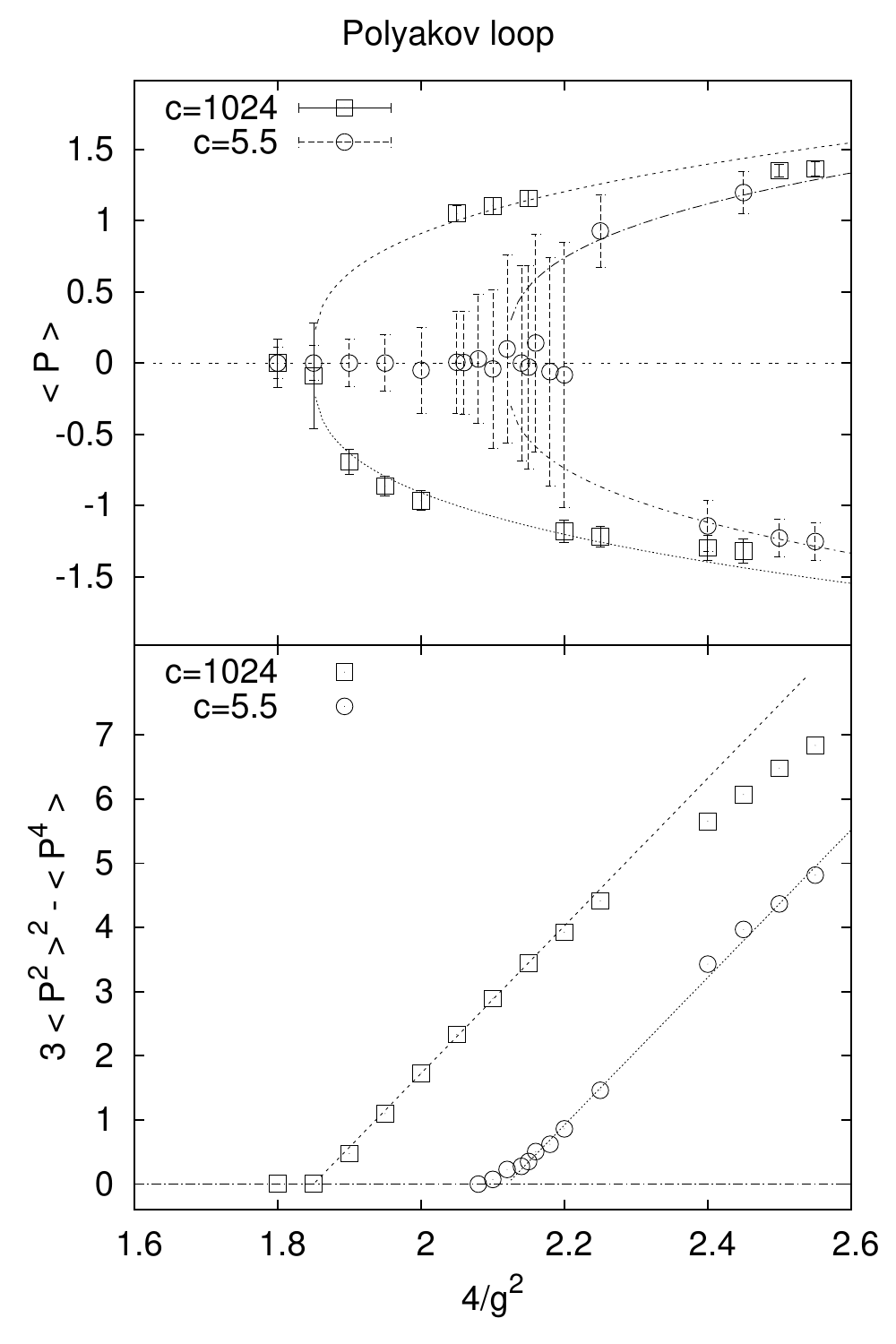}
\caption{\label{FIG1:POL}
 Polyakov-loop expectation values and fourth order cumulants on $10^3 \times 2$
 lattices at $c=5.5$ (circles) and at $c=1024.0$ (squares). 
 The critical coupling, $x=4/g^2_c$, is estimated by fitting an 
 $\sim(x-x_c)^{1/3}$ function to the $< P >$ values and  
  with a linear fit to the smaller nonzero values in case of the cumulants.  
}
\end{figure}

The deconfining transition can also be studied within the 
superstatistical framework by investigating the behavior of the Polyakov 
loop. Results on this are illustrated in Fig.\ref{FIG1:POL} 
for $c = 5.5$ (a realistic value from $p_T$ spectra) and for $c = 1024$
(which represents effectively the traditional, Gibbs limit). 
The Polyakov loop expectation values and the fourth order cumulants have been
calculated and the critical couplings have been determined using a 
functional fit to the data. At $c=1024$ the result is 
$x_c = 4/g_c^2 = 1.85$ (for both quantities) which is in a good agreement 
with the known result for the SU(2) LGT on $N_t=2$ lattices 
\cite{Fingberg,Velytsky}.
At $c=5.5$ one obtains $x_c=2.12$ with the same type of fits (again, 
consistently for the Polyakov loops and the cumulants as well) which
means that the transition temperature moves towards higher values
in that case. We are interested in the amount of the change in the 
temperature values. Based on \cite{Velytsky} the critical 
couplings can be translated to temperatures, and if we denote the 
critical temperature for the canonical Gibbs ensemble with $T_c$,
for $c=1024$ the transition occurs (trivially) at $T_c$ and
for $c=5.5$ one gets the value $T \approx 1.3 T_c$.

This result, taken at face value, suggests a considerable increase of the 
transition temperature, supposedly due to non-equilibrium effects.
However, one could argue that during the simulation $<\theta> =1$,
i.e. it is the inverse temperature, which has the mean value 
of $1/T$. Due to the properties of the Gamma distribution, the 
mean of the temperature itself is determined by 
$ <1/\theta> = c/(c-1) \approx 1.22 $. Nevertheless, the dynamical
effect is definitely larger than the trivial statistical factor of 1.22
obtained by this argument.

One should make a further remark here. In the above superstatistical
Monte Carlo method the anisotropy $\theta$ is actually the bare 
anisotropy. This parameter should be regarded as an additional 
coupling (or, one can imagine having different coupling parameters
in spacelike and timelike directions on the lattice) and 
it should be renormalized accordingly in order to obtain 
physical results in the simulation. In principle this renormalization
can be incorporated in the above algorithm if the functional form
between the bare and renormalized anisotropies is known. Then an
additional step is required in the update sweep: after generating 
the actual anisotropy parameter $\theta$ according to Gamma distribution 
(which should be called the physical anisotropy) one should calculate 
the bare anisotropy $\theta_0$ using this functional form and perform 
the update with $\theta_0$ (which, in that case is not really 
Gamma distributed any more). 

For this modification one would
actually need to know $\theta_0$ as a function of $\theta$ in 
the whole $(0, \infty)$ interval to cover the whole range of the 
Gamma distribution. As anisotropic lattices are also used
in the usual simulations, this function has been determined 
\cite{Klassen}, although only for $\theta \le 1$. The reason is 
practical: anisotropic lattices are used mainly for simulating
systems at finite (high) temperature and for that purpose 
$\theta \le 1$ suffices. As a consequence, for renormalizing the
superstatistical lattice gauge theory, first the above function 
should be determined for the region $\theta > 1$. That could be 
non trivial, although some relevant physical quantities do not 
depend strongly on the temperature below the deconfining transition.

Because of these difficulties, renormalization is not considered
here. Clearly, the procedure would influence the value of the 
transition temperature. Nevertheless, we think that the qualitative
results obtained from the superstatistical Monte Carlo simulation 
are still reliable. Therefore it is safe to conclude
that experiments aiming at producing quark matter under circumstances 
characteristic to high energy collisions should  consider the possibility 
of an up to about $30$ per cent higher $T_c$ then predicted by traditional 
Monte Carlo lattice calculations.
(Let's mention that similar effects are reported -- although in opposite 
direction -- based on completely different studies: including more 
and more quark fields in the QCD Lagrangian \cite{Lombardo15}, or applying 
strong external magnetic fields \cite{Bonati16}, the critical temperature 
seem to decrease.)




\section{Conclusion and Open Problems}


In conclusion we gave a review of several ''exotic'' methods dealing with the QCD deconfinement
transition from hadron to quark and gluon dominated matter and back, picking out both phenomenological
model approaches and computation-technique developments.
Our main line of thoughts of dwelling into questions of the complexity of the structure
of quark-gluon plasma (QGP) was accompanied by the playful experimentation with a possible
extension of the Boltzmann-Gibbs canonical view in thermodynamics for non-equilibrium
phenomena and finite reservoir effects.

Keeping as a main subject the exploration of the well-established high-temperature
field theory approach, we contrasted a number of phenomenological approaches, also as 
speculative physical models, with the most important findings. There is no sudden
switch from the hadronic world to an {\em ideal} quark-gluon plasma.
The complex universe of a QGP in the range $T_c - 4T_c$, also called an sQGP by some,
is full with beautiful physical effects. These effects go far beyond an effective
quasiparticle mass, which is more or less proportional to the temperature.
The increasing thermal width and the occurrence of continuum parts in the physical spectral
function have competing effects on the total pressure. Their balance results in an
equation of state describing a matter softer than the radiative Stefan-Boltzmann gas
with massless constituents.

This phenomenon can be interpreted within a string-like interaction picture considering
the (color-)density dependent contribution to the free energy density. Also an assumption
of emerging and melting quasiparticle peaks in the spectral function, as well as a higher
correlation effect on generated mass terms agrees with both with leading order pressure
corrections in perturbative QCD and with numerically obtained lattice Monte Carlo data.

Beyond the equilibrium view, very few things can be done in exact quantum field theory.
The linear response approximation allows us to wring out some information on transport
properties near the thermal equilibrium; most famous being the shear viscosity coefficient.
Here also some traps occur in thinking, we have tried to point out some of them in this review.
On the other hand we have introduced arguments in favor of considering non-Gibbsean
weighting of configurations as if we were dealing with nonlinear (non-Glauber) quantum
coherent states of radiation or -- with a very similar effect -- with an unsteady environment
resembling an Euler-Gamma distributed inverse temperature, as superstatistics.

Certainly we have left a number of problems undisclosed. Just to list a few
we formulate questions for future consideration, emerging from high energy experimental results:
Is most of entropy produced initially or during a fast hadronization at the end of these reactions?
Is there any real thermalization (even if no sharp temperature is possible in an event--averaged
ensemble of measured data) or are all temperature effects just illusion, reflecting
barely an Unruh-like temperature (acceleration based, with no heat container)?
How to produce dynamical states closely but not exactly reminding to thermal equilibrium states 
without ever reaching an equilibrium?

We close this short review with the hope that the Reader could find as much enjoyment
in passing these ideas through his/her mind, as the Authors definitely did.

\begin{acknowledgement}
This work has been supported by the Hungarian National
National Research, Development and Innovation Office (NKFIH)
under the contract numbers OTKA 104260 and 104282.
\end{acknowledgement}
%

\end{document}